\documentclass[5p,twocolumn,10pt,times]{elsarticle}
\usepackage{amsmath}
\usepackage{hyperref}
\usepackage{wrapfig}
\usepackage{soul}
\usepackage{array}

\newtheorem{proposition}{Proposition}

\usepackage{color}

\newcommand{\revision}[2]{\textcolor[rgb]{0.00,0.00,0.00}{#2}}

\begin{document}
\baselineskip11pt

\begin{frontmatter}

\title{Vector Field Based Volume Peeling for Multi-Axis Machining}

\author{Neelotpal Dutta\textsuperscript{*}}
\author{Tianyu Zhang\textsuperscript{*}}
\author{Guoxin Fang}
\author{Ismail E. Yigit}
\author{Charlie C.L. Wang\textsuperscript{**}}
\cortext[cor1]{Joint first authors}
\cortext[cor2]{Corresponding authors. E-mail: changling.wang@manchester.ac.uk}

\address{Digital Manufacturing Laboratory, School of Engineering, The University of Manchester, United Kingdom}


\begin{abstract} 
This paper presents an easy-to-control volume peeling method for multi-axis machining based on the computation taken on vector fields. The current scalar field based methods are not flexible and the vector-field based methods do not guarantee the satisfaction of the constraints in the final results. We first conduct an optimization formulation to compute an initial vector field that is well aligned with those anchor vectors specified by users according to different manufacturing requirements. The vector field is further optimized to be an irrotational field so that it can be completely realized by a scalar field's gradients. Iso-surfaces of the scalar field will be employed as the layers of working surfaces for multi-axis volume peeling in the rough machining. Algorithms are also developed to remove and process singularities of the fields. Our method has been tested on a variety of models and verified by physical experimental machining. 
\end{abstract}


\end{frontmatter}


\section{Introduction}

    Process planning is one of the most important stages in manufacturing. In the context of CNC milling, once the types of machine and material have been chosen, the process planning involves the decisions like orientation selection, volume decomposition, and tool-path generation \cite{ren2010integrated}. The factors to be considered during process planning include tool accessibility (i.e., collision avoidance), machining efficiency, surface quality, damage to the cutting tool, etc. These factors are in turn also influenced by the hardware and the material choices taken in the earlier step. An important consideration during milling or machining is that of \textit{manufacturability}. While the term manufacturability can include many definitions, we consider the following criteria as the primary requirements in this work
\begin{enumerate}
    \item the presence of collision-free directions of the tool for every point on the planned working surfaces; 

    \item the absence of any unconnected floating volume of material during the machining process.
\end{enumerate}
These are major problems to be resolved when giving an arbitrary volume decomposition that is not manufacturable. 

\begin{figure}[t]
 \centering 
\includegraphics[width=\linewidth]{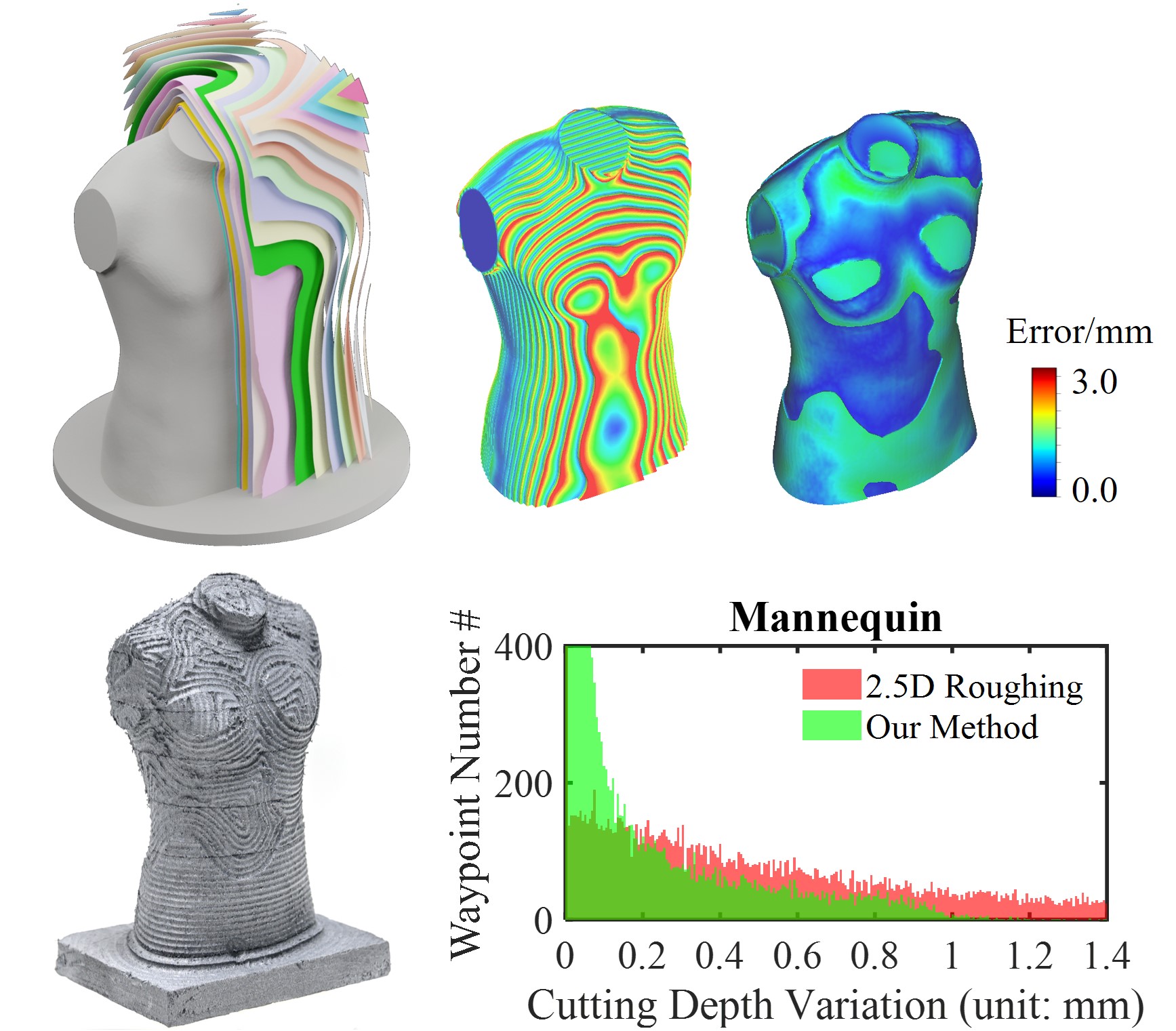}
\put(-252,105){\small \color{black}(a)}
\put(-165,105){\small \color{black}(b1)}
\put(-102,105){\small \color{black}(b2)}
\put(-252,3){\small \color{black}(c)}
\put(-160,3){\small \color{black}(d)}
\caption{Process of roughing is formulated as a problem to compute optimized layers that indicate the volume peeling procedure to remove materials between layers gradually: (a) curved layers for volume peeling, (b) the shape approximation errors w.r.t. the target model on the 2.5D roughing result (b1) vs. the errors on our result (b2), (c) the physical result of roughing by our method, and (d) the histogram to compare the cutting depth variations during finishing from the model obtained by 2.5D roughing (b1) vs. the model of our method (b2). 
}\label{fig:layerBasedProcPlan}
\end{figure}
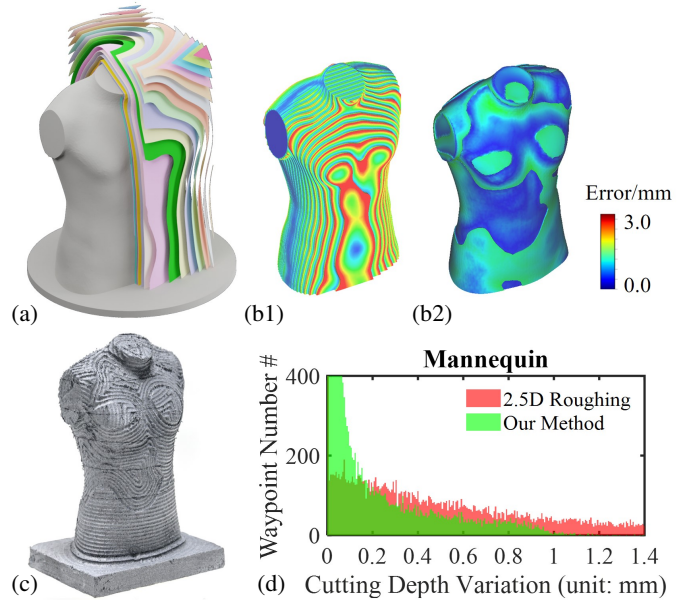

The process of milling can be broadly divided into two steps: rough milling (or called \textit{roughing}) and finish milling (or named as \textit{finishing}). The roughing step precedes the finishing step and is responsible to bring the stock of material from its initial shape to a shape as close as the target model by removing materials. As this step needs to remove bulks of the stock material, it is usually termed as a volumetric operation. Machining at this stage aims to remove \textit{as much material as possible in less number of feasible operations}. 
The other stage, finishing, involves the fine removal of material to introduce all the designed features of the target model as well as to reach the desired surface quality within a given tolerance. This paper focuses on the roughing step -- i.e., planning for volumetric operations (see Fig.\ref{fig:layerBasedProcPlan}). 

\subsection{Problem of Volume Peeling}
Due to the ease of computation, the volume of stock material is usually \textit{peeled} into planar layers in many computer-aided manufacturing (CAM) software systems. However, such an approach leads to the problem of staircases and cannot effectively remove material in the shadow regions according to the fixed orientation of the machine tool. These drawbacks increase the total machining time and also lead to the problems of large cutting force fluctuation during finishing (Ref.~\cite{lauwers2006five, li2020voxel, he2021geodesic}).

A better strategy of roughing is expected to achieve the highest possible removal rate while obtaining a `near-net' shape at the end of roughing so that only a minor effort of finishing is needed to complete the geometric details. To achieve this, non-planar peeling methods were introduced and had to be physically realized on a multi-axis machine. The multi-axis roughing approaches can be classified into the morphing methods \cite{lauwers2006five,huang2013unified,chen2018variable}, the offsetting method \cite{zhu2007rough}, and the field based methods \cite{li2020voxel, he2021geodesic}. The plans of roughing (i.e., layers of working surfaces) generated by these methods are not easy-to-control to accommodate complicated features or extra constraints. They also face problems like self-intersection (e.g., \cite{zhu2007rough}) and non-uniform depth-of-cut (e.g., \cite{lauwers2006five}). A new method is needed.

In this paper, we propose a vector field based method of volume peeling for multi-axis machining, which provides an easy-to-control tool to generate the manufacturing layers for roughing. Although the vector field has been widely used in the area of computer numerical controlled machining to guide the motion of tool on the surface (e.g., \cite{kim2002toolpath, my20055, xu2014five, bo2019initialization}), it has not been explicitly used in volume deposition for roughing as of now. The fields in these existing approaches were mainly designed to achieve results like maximized cutting width (therefore minimized machining time) and gouging avoidance. On the other aspect, scalar field based methods like the geodesic field based decomposition presented in \cite{he2021geodesic}, which is less flexible to provide the `controllability' as what is offered by the vector field based approach in this paper. Our method is mainly motivated by the fact that simple local constraints in the vector field can translate to a number of desirable properties in the resultant shape or deformation process \cite{von2006vector, yu2004mesh}. We explore the feasibility of using a vector field based method to develop an easy-to-control tool for volume peeling. Being interactive, it allows a user to change the shape of the generated layers of manufacturing surfaces through simple and local instructions.
%

\subsection{Related Work} \label{subsecRelatedWork}
\subsubsection{Process Planning for Five-Axis Rough Milling}
The automation of process planning and tool-path generation for machining has been extensively studied for several years \cite{pi1998grind, li1993automatic, 10.1145/1499949.1500131, halevi1980development, subrahmanyam1995overview, liang2021review, kukreja2023efficient}. However, most of the research on tool-path generation has been focused on finishing operations due to their impact on the surface quality of the output. There has been comparatively limited investigation into rough milling processes, despite rough machining accounting for a significant portion of machining time \cite{li1994optimal, heo2008efficient}. Planar approaches, such as the Z-level method, are widely used for material removal during rough machining \cite{yuen1987octree, tseng1999machining}. Li et al. \cite{li1994optimal} have analyzed various tool-path patterns for 2.5D planar layers of rough machining, while Balasubramaniam \cite{balasubramaniam2001automatic} presents a method for automatically planning 5-axis tool-paths to improve accessibility. \revision{}{Joneja et al. \cite{joneja2003greedy} have presented a method to reduce the rough milling time by heuristically using multiple tools, where larger tools are used to remove larger amount of material initially.} However, these methods still involve the planar decomposition of the volume, resulting in significant staircase effects that require extra processing and unwanted dynamic effects during finishing \cite{lauwers2006five,li2020voxel, he2021geodesic}.

To address these issues, Lauwers and Lefebvre \cite{lauwers2006five} proposed a morphing-based layer generation strategy for 5-axis roughing, which has also been explored in a similar strategy in \cite{huang2013unified}. Additionally, the algorithm in \cite{zhu2007rough} involves using the offsetting method to produce non-planar layers for roughing. Multiple studies have investigated the 5-axis rough machining of impellers \cite{heo2008efficient, young2004five}, but these methods have limited applicability. More recently, scalar-field propagation-based layer generation methods have been explored \cite{he2021geodesic, li2020voxel}, although they lack flexibility to accommodate other requirements.

\subsubsection{Vector Field for Manufacturing-Planning}
The use of vector fields (or direction fields) in various manufacturing applications has been widely investigated. For instance, in \cite{kim2002toolpath, xu2014five, chiou2002machining}, fields of optimal cutting directions are used to generate toolpath on surfaces, leading to reduced machining time and improved surface quality. Similarly, in \cite{bo2019initialization}, a vector field is generated on a general free-form surface to approximate the given surface by identifying the orientations of the milling tool. Sun et al. \cite{sun2018iso} use an iso-planar feed direction to create a smooth $G^1$ continuous toolpath for surface machining. Another class of methods uses vector fields to partition the surface into smaller patches, resulting in higher efficiency. For example, in \cite{li2021partition}, a surface contact (vector) field is used to partition the surface into patches of similar tool orientation and feed direction. A similar strategy is used by \cite{my20055}. However, these methods are surface-based and do not consider the volume-based process of rough machining. 

In the field of multi-axis additive manufacturing, vector fields have been used to decompose the volume into several layers. For example,  Fang et al. \cite{fang2020reinforced} have introduced a stress field-aligned vector field to slice the model into curved layers. In the method presented by Li et al. \cite{li2022vector}, the vector fields are designed to satisfy multiple manufacturing constraints by manipulating the vector directions at different constraint points. However, these methods do not consider the integrability of the vector fields.
In \cite{mahdavi2020vdac}, an accessibility field is used to segment the volume of the material to be machined. But this method uses a 2.5D tool-path to remove material from each volume segment. The scalar field based method presented in \cite{he2021geodesic} creates a vector field as an intermediate stage but the vector field is not exploited to introduce more degree of freedom in the process planning.

\subsubsection{Shape Modeling by Vector Field}
The use of vector field based methods extends beyond manufacturing applications. In recent years, these methods have also been used for shape editing and denoising. For instance, in the algorithm presented by Yu et al. \cite{yu2004mesh}, a gradient field and Poisson equation are used to perform interactive mesh deformation, merging, and denoising. Similarly, Liao et al. \cite{liao2009gradient} use a gradient-based method for the deformation of tetrahedral mesh. A vector field is employed in \cite{von2006vector} to obtain intersection-free mesh deformation. Another application of vector fields is found in \cite{kazhdan2006poisson}, where the solution of a Poisson equation is used to reconstruct surfaces from point data through a gradient field. More generally, quaternion fields (as generalized vector fields) are used by Zhang et al. \cite{zhang2022s3} to perform a flexible multi-axis slicing, allowing sliced layers to easily take the desired shape to improve functions such as stress-reinforcement, support-free printing, and surface finish. 

\begin{figure*}[t]
\centering 
\includegraphics[width=0.98\linewidth]{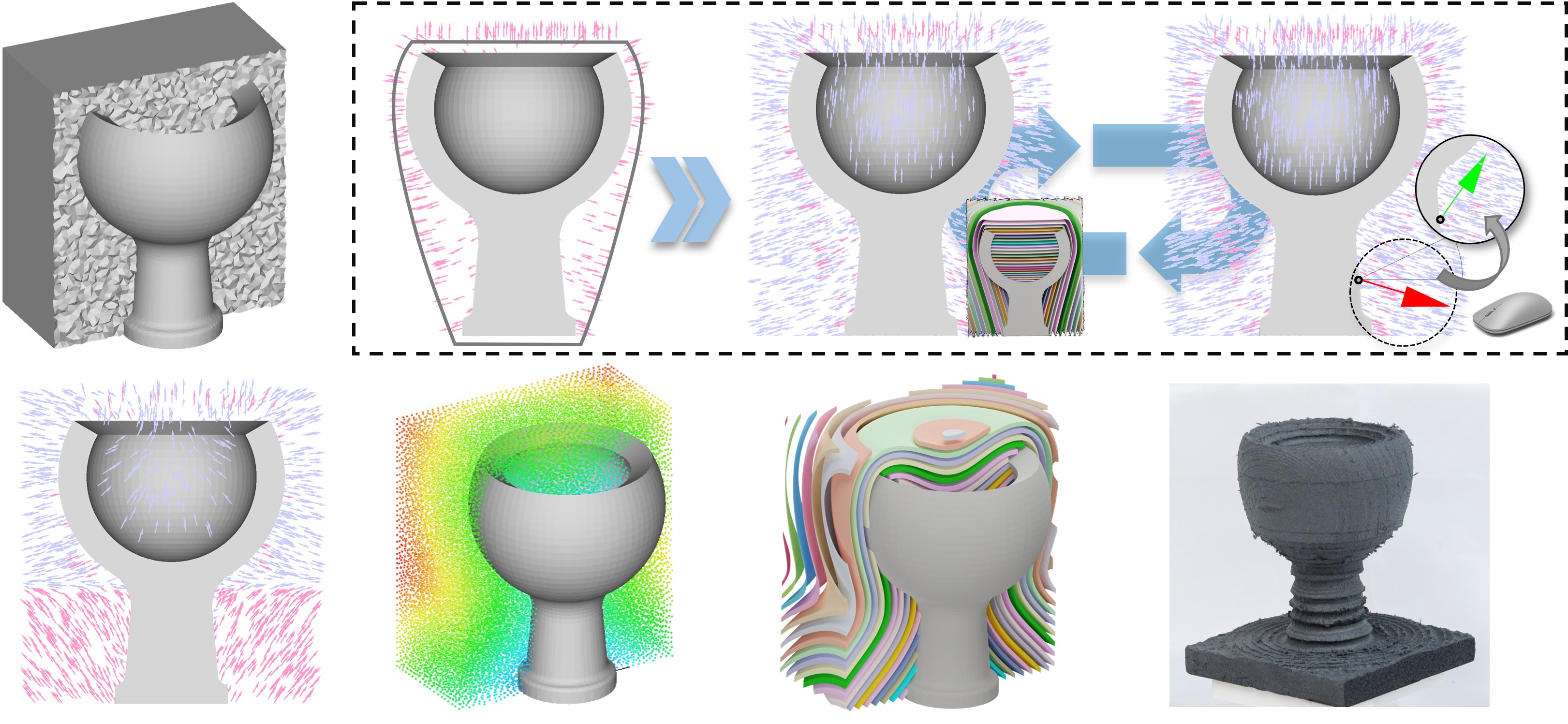}
\put(-526,128){\small \color{black}(a)}
\put(-402,128){\small \color{black}(b)}
\put(-526,6){\small \color{black}(c)}
\put(-402,6){\small \color{black}(d)}
\put(-272,6){\small \color{black}(e)}
\put(-145,6){\small \color{black}(f)}
\put(-470,177){\small \color{black}$\mathcal{H}$}
\put(-519,236){\small \color{black}$\mathcal{M}$}
\put(-450,230){\small \color{black}$\Omega$}
\vspace{-5pt}
\caption{Diagram for showing the pipeline of our controllable working surface generation method. (a) The working domain $\Omega$ of removable material defined by the difference between the stock $\mathcal{M}$ and the model $\mathcal{H}$ is represented by a tetrahedral mesh in our approach. (b) By using anchor vectors from the surface normals of $\mathcal{H}$'s convex hull (i.e., pink arrows in the left), where the vector field $\mathbf{v}(\mathbf{x})$ can be generated by smoothly interpolating the anchor vectors  -- see the newly generated blue arrows in $\Omega$ (middle). The vector field can be easily edited by selecting a set of vectors (e.g., the red one shown in the right sub-figure of (b)) to change their orientations (e.g., adjusting to the green one). (c) The new vector field generated according to this change can be computed by an optimization framework. (d) After optimizing the vector field into an irrotational one, a scalar field $g(\mathbf{x})$ can be determined by solving the Poisson's equation (i.e., Eq.(\ref{eq:v_div})). (e) Iso-surfaces are extracted from $g(\mathbf{x})$ as the working surface layers $\{\mathcal{L}_i\}$. (f) The physical roughing result by using these working surface layers $\{\mathcal{L}_i\}$.
}\label{fig:pipeline}
\end{figure*}
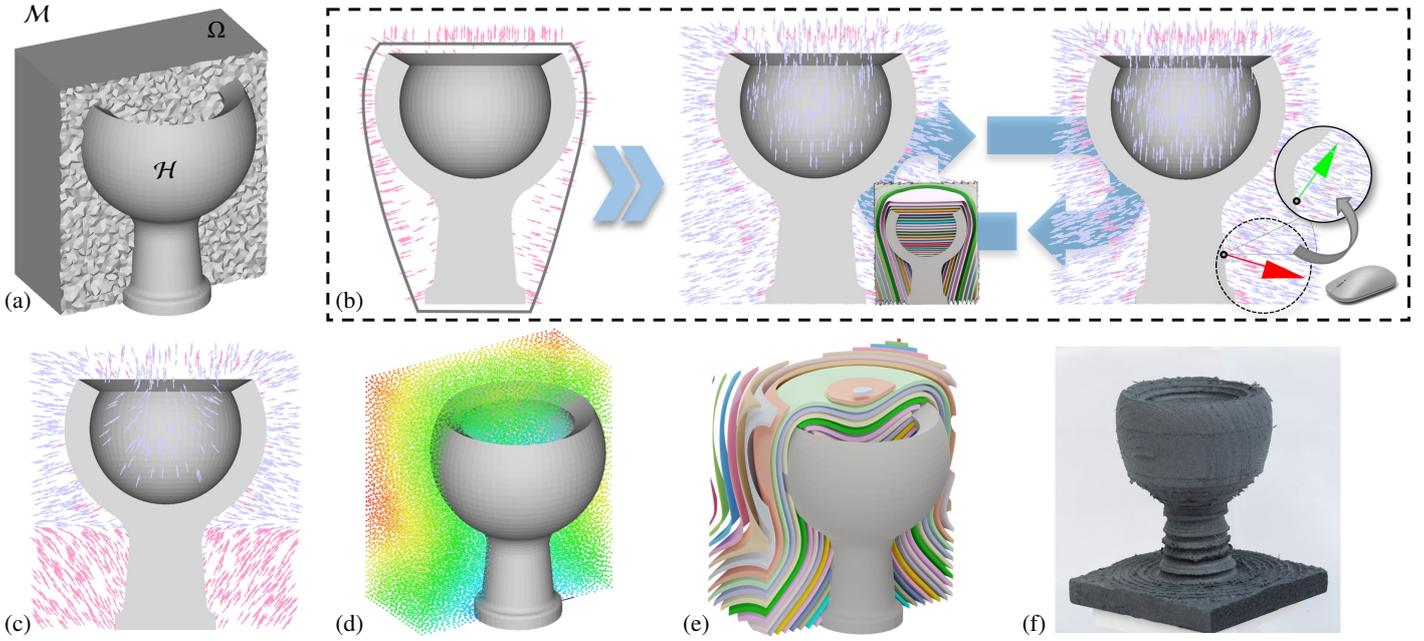

\subsection{Contribution}\label{subsecContribution}
We propose a vector field based method of controllable volume peeling for the roughing plan of multi-axis machining. The following contributions are made in our work: 
\begin{itemize}
\item With the help of vector fields, a new computational pipeline is introduced that can easily control the working surface layers generated for the volume peeling process to be used in multi-axis rough machining.

\item The analysis and solution for the singularity of vector / scalar fields have been developed, which is important to ensure the manufacturability of the results obtained from volume peeling.

\item An interactive interface has been developed to provide easy-to-use tools for controlling the layers of volume peeling therefore also the machining process.
\end{itemize}
Our method has been successfully tested on a variety of models. Physical experiments have also been conducted to verify its performance.

\section{Formulation and Analysis}
    \subsection{Process Planning}\label{subsecProcFormulation}
Considering a model $\mathcal{H}$ that is to be manufactured from the stock of material $\mathcal{M}$ by removing the volume $\Omega = (\mathcal{M} - \mathcal{H})$, the problem of process planning for roughing can be formulated as to determine the gradual order of  material removal from the volume $\Omega$. Following the strategy of conformal volume peeling, we wish to decompose $\Omega$ into stacks of surfaces $\{\mathcal{L}_i\}_{i=1,2,\cdots,n}$ so that materials can be removed layer by layer (i.e., the $i$-th layer indicates the material between $\mathcal{L}_i$ and $\mathcal{L}_{i+1}$). This has been illustrated in Fig.\ref{fig:layerBasedProcPlan}.

We propose to compute this decomposition by extracting the iso-surfaces of a scalar field $g(\mathbf{x})$ as $\{\mathcal{L}_i\}_{i=1,2,\cdots,n}$, where
\begin{equation}
    \mathcal{L}_i = \{\mathbf{x} \; | \; \forall \mathbf{x} \in \mathbb{R}^3, g(\mathbf{x}) = c_i \}
\end{equation}
with $c_i$ being the iso-value for the iso-surface $\mathcal{L}_i$. \revision{}{Note that each $\mathcal{L}_i$ can be a union of disconnected iso-surfaces corresponding to the same iso-value.} The scalar field $g(\mathbf{x})$ can be flexibly constructed and controlled via a vector field $\mathbf{v}(\mathbf{x})$ in $\Omega$ such that $\nabla g(\mathbf{x}) =  \mathbf{v}(\mathbf{x})$ ($\forall \mathbf{x} \in \Omega$). Here $\nabla$ represents the gradient of a scalar field. Considering a point $\mathbf{x} \in \Omega$ that lies on the iso-surface as $\mathbf{x} \in \mathcal{L}_i$, the surface normal of $\mathcal{L}_i$ at $\mathbf{x}$ is directly indicated by the vector $\mathbf{v}(\mathbf{x})$ when $\nabla g(\mathbf{x}) =  \mathbf{v(\mathbf{x})}$ is \textit{strictly satisfied}. This also means that we can `transmit' a few locally defined orientations to the shape control of surface layers $\{\mathcal{L}_i\}$ -- therefore also the volume deposition plan for multi-axis machining. This provides an easy-to-control tool for process planning (see Fig.\ref{fig:pipeline} for an illustration).

The method proposed in this paper will enable the function of generating the vector field $\mathbf{v}(\mathbf{x})$ according to the target model shape $\mathcal{H}$, the stock material shape $\mathcal{M}$, and the user-specified control vectors $\{\mathbf{a}_j\}_{j=1,2,\cdots,m}$ defined on the locations $\{\mathbf{x}_j\}_{j=1,2,\cdots,m}$ (with $\mathbf{x}_j \in \Omega$).
To ensure the strict satisfaction of $\nabla g =  \mathbf{v}$ to realize a direct control of volume peeling, the \textit{irrotational} property needs to be imposed on $\mathbf{v}(\mathbf{x})$. \revision{}{Additional requirements about singular cases are introduced to ensure that the vector field represents a well-defined gradient inside $\Omega$ (see Sec.~\ref{SecVFImprove}).}

Moreover, the final objective is to ensure \textit{manufacturability} during the whole process. As defined above, this means the conditions of accessibility and none-floating-volume are satisfied. A strong accessibility requires the points to be accessible along the planned direction and the trajectory moving from point to point to be collision-free. And a weak condition can be stated as the presence of at least one accessible direction for every point on the toolpath. Our aim is to ensure strong accessibility condition. 


\subsection{Irrotational Vector Field}\label{subsecVecFieldAnalysisIrrotational}
Given a vector field $\mathbf{v}: \mathbb{R}^3 \rightarrow \mathbb{R}^3$ defined in a \textit{simply connected domain}, it can be decomposed as follows according to the Helmholtz-Hodge decomposition \cite{Harsh2013the}:
\begin{equation}
    \mathbf{v} = \nabla g + \nabla \times \mathbf{r} 
    \label{eq:HodgeDecomp2}
\end{equation}
Here, the first term $\nabla g$ is the gradient of a scalar field $g(\mathbf{x})$ (known as the \textit{scalar potential}), and the second term $\nabla\times \mathbf{r}$ represents the curl of another vector field $\mathbf{r}(\mathbf{x})$ which is known as the \textit{vector potential}. 
Moreover, for a bounded domain, there is a requirement of the third vector field term which has both zero divergence and zero curl (named as the harmonic vector field). This however can be represented as the additional part included in either the curl of a vector field or the gradient of a scalar field. In our case, we assume that it is included in the gradient term.

From the properties of the gradient, the divergence and the curl, Eq.\eqref{eq:HodgeDecomp2} can be further processed as:
\begin{equation}
    \nabla \cdot \mathbf{v} = \nabla \cdot \nabla g + \nabla \cdot \nabla  \times \mathbf{r} = \Delta g,
    \label{eq:v_div}
\end{equation}
based on the fact that $\nabla \cdot \nabla  \times \mathbf{r} = 0$. 
This is actually a Poisson's equation. Solving Eq.\eqref{eq:v_div} will give a scalar field $g(\mathbf{x})$, the gradient of which is the best approximation of $\mathbf{v}(\mathbf{x})$. However, $\nabla g(\mathbf{x})$ may still have a certain level of deviation from the control anchor vectors $\{\mathbf{a}_j\}$ because that $\nabla g(\mathbf{x})$ only represent the irrotational part of $\mathbf{v}(\mathbf{x})$ -- i.e., the first term of the Helmholtz-Hodge decomposition. 

%

\begin{figure}[t]
\centering 
\includegraphics[width=\linewidth]{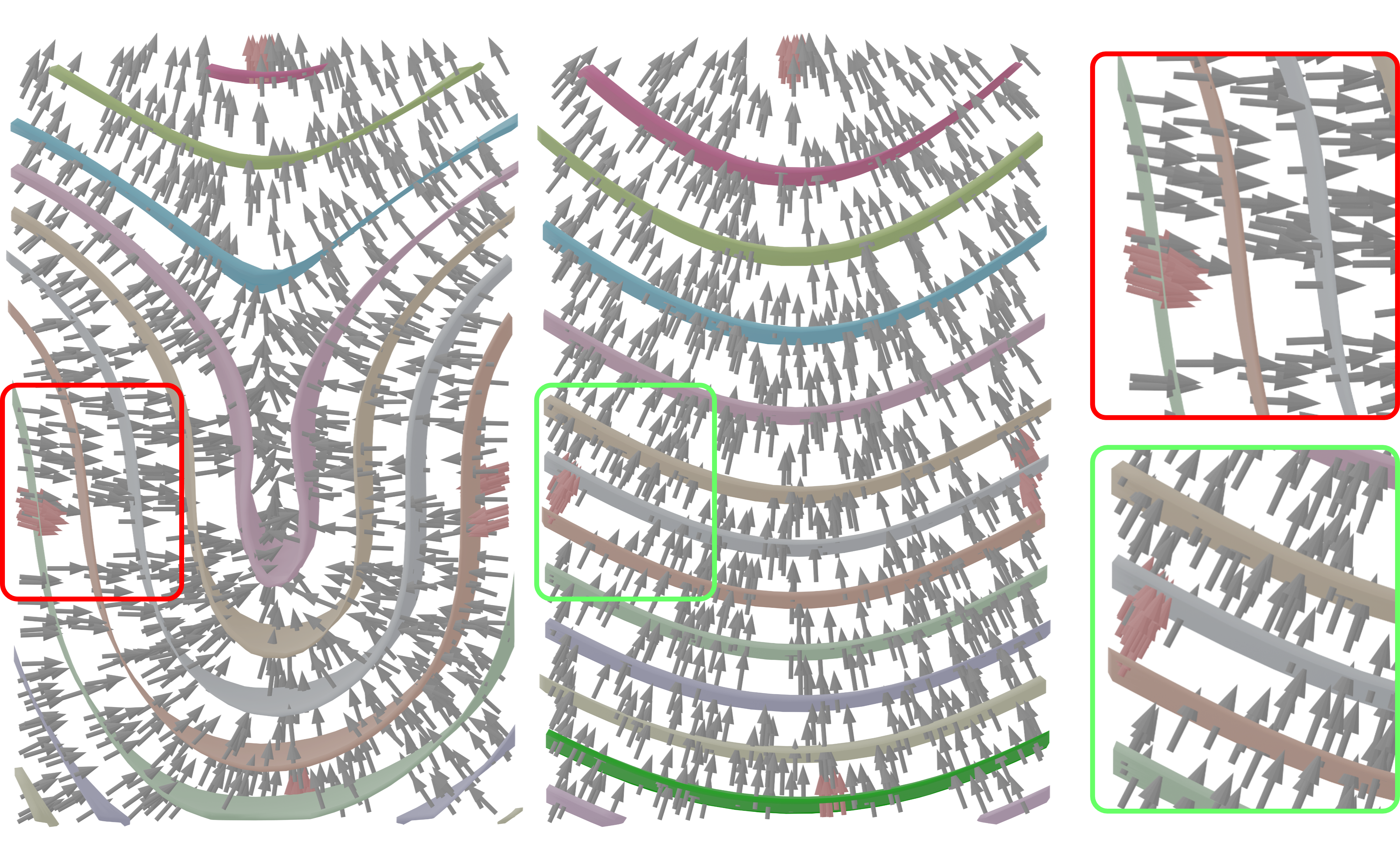}
\put(-210,-4){\small \color{black}(a)}
\put(-115,-4){\small \color{black}(b)}
\put(-32,-4){\small \color{black}(c)}
\caption{Examples to highlight the importance of curl-free property for the agreement between a vector field $\mathbf{v} (\mathbf{x})$ and the gradient of a corresponding scalar field $g (\mathbf{x})$ determined by the Poisson's equation (i.e., Eq.\eqref{eq:v_div}). The example field in (a) has $I_{rot}=4.50 \times 10^{-2}$; differently, the example field (b) has $I_{rot}=3.59\times10^{-4}$ -- both vector fields have been normalized as $\|\mathbf{v}\|=1$. From the zoom views in (c), it can be easily observed that there are large variations between $\nabla g (\mathbf{x})$ (as the normal vectors of $g (\mathbf{x})$'s iso-surfaces) and the vector field $\mathbf{v} (\mathbf{x})$ defined in (a) while there is nearly no variation for the field defined in (b). 
}\label{fig:ProblemFormulation}
\vspace{-10pt}
\end{figure}

Different from directly approximating the vector field $\mathbf{v}(\mathbf{x})$ by the gradients of a scalar field via energy minimization (ref.~\cite{fang2020reinforced}), we propose to compute a vector field $\Tilde{\mathbf{v}}(\mathbf{x})$ that only contains an irrotational field while preserving the field $\mathbf{v}(\mathbf{x})$ at some critical regions (like areas prone to singularity that will be discussed later) and relaxing at others. This is important to ensure that the generated scalar field does not have unwanted distortions (see Fig.~\ref{fig:ProblemFormulation}(a) as an example with significant rotational component). The control will be realized via the imposed anchor vectors. 
\begin{proposition}
To generate volume peeling layers via a scalar field $g(\mathbf{x})$ with $\nabla g(\mathbf{x}) = \mathbf{v}(\mathbf{x})$ \textit{strictly satisfied}, $\mathbf{v}(\mathbf{x})$ should be an irrotational vector field that has zero curl for its vector potential, i.e., $\nabla \times \mathbf{r} = 0$. \label{prop:curl}
\end{proposition}
Given a vector field $\mathbf{v}(\mathbf{x})$, we define a metric as follows to evaluate its likelihood to be an irrotational vector field
\begin{equation}\label{eq:likelihood}
    I_{rot}= \frac{1}{| \Omega |} \int_{\Omega} \|\nabla \times \mathbf{r} \|^2d \Omega
\end{equation}
with $| \Omega |$ being the volume of the domain and $\nabla \times \mathbf{r}$ being the second term of $\mathbf{v}(\mathbf{x})$'s Helmholtz-Hodge decomposition. By Eq.\eqref{eq:HodgeDecomp2}, we know that this metric can also be evaluated by
\begin{equation}\label{eq:likelihood2}
    I_{rot}= \frac{1}{| \Omega |} \int_{\Omega} \|\mathbf{v} - \nabla g \|^2d \Omega.
\end{equation}
See Fig.\ref{fig:ProblemFormulation} for two examples with different values of $I_{rot}$.

The solution of Poisson's equation on an irrotational field $\Tilde{\mathbf{v}}(\mathbf{x})$ should give us a scalar field $g(\mathbf{x})$, the gradient of which is exactly the field $\Tilde{\mathbf{v}}(\mathbf{x})$. When the irrotational vector field $\Tilde{\mathbf{v}}(\mathbf{x})$ interpolates the control anchor vectors $\{\mathbf{a}_j\}$, we can then generate a scalar field $g(\mathbf{x})$ with these anchor vectors strictly satisfied by $\nabla g(\mathbf{x})$. As a result, $g(\mathbf{x})$'s iso-surfaces as the layers of volume peeling are well controlled by the anchor vectors. This fact has been illustrated in Fig.~\ref{fig:ProblemFormulation}. The algorithm to improve the likelihood of an irrotational vector field will be presented in Sec.~\ref{subSecCurlRemov}.

\subsection{Singularity of Vector Field} \label{subsecVecFieldAnalysisCurl}
For a vector field $\mathbf{v}(\mathbf{x})$, it can have a vector distribution with sources or sinks in the domain. More generally, sources and sinks are point singularity with vanished vector magnitude. In other words, at a singular point $\mathbf{p}$, it has $\mathbf{v}(\mathbf{p}) = 0$ with the direction of vector vanished. In our case, the volume peeling layers are generated from the iso-surfaces of a scalar field $g(\mathbf{x})$ that is determined from $\mathbf{v}(\mathbf{x})$ by solving the Poisson's equation (i.e., Eq.\eqref{eq:v_div}). These point singularities will directly lead to local minima or maxima on $g(\mathbf{x})$ -- i.e., isolated regions will be formed during the process of volume peeling when these singularities not are located on the boundary of computational domains. As will be explained later, these isolated regions will form working surface layers that cannot be accessed in a collision-free way.
\begin{proposition}
    If a vector field is designed with the aim of using it to generate layers of volume peeling for machining, it should have no source or sink in the interior of the domain which requires $\nabla \cdot \mathbf{v} (\mathbf{x}) = 0$  ($\forall \mathbf{x} \in \Omega - \partial \Omega$). 
    \label{prop:singularity}
\end{proposition}
Satisfying $\nabla \cdot \mathbf{v} (\mathbf{x}) = 0$ on an irrotational vector field $\mathbf{v} (\mathbf{x})$ gives $\Delta g = 0$ (by Eq.\eqref{eq:v_div}), which means that $g(\mathbf{x})$ needs to be a harmonic field without interior local minimum/maximum. Sec.~\ref{subSecSingularityResol} will present an algorithm to process / remove the singularities.

\section{Field Generation and Layer Extraction}
\label{secFieldGen}
    \subsection{Primitives}
In our method, we represent the volume $\Omega$ to be machined by a tetrahedral mesh (i.e., $\Omega = \{\mathcal{T}, \mathcal{F}, \mathcal{E}, \mathcal{V}\}$) that consists of tetrahedra ($\mathcal{T}$), faces ($\mathcal{F}$), edges($\mathcal{E}$) and vertices ($\mathcal{V}$). The element-based representation is employed for the vector field $\mathbf{v}(\mathbf{x})$ -- i.e., the value of $\mathbf{v}(\mathbf{x})$ is a constant inside each tetrahedral element. The scalar field $g(\mathbf{x})$ is represented in the piece-wise linear manner by storing scalar values on the vertices. Specifically, we have 
\begin{equation}\label{eq:scalar_interpolation}
    g(\mathbf{x}) = \sum_{v_i \in \mathcal{T}_k \,:\, \mathbf{x} \in \mathcal{T}_k} h_i(\mathbf{x}) g_i,
\end{equation}
where $h_i$ is a linear hat function such that $h_i$ is unity at the $i^{th}$ vertex ($\mathcal{V}_i$) and zero at every other vertex.  $g_i$ is the scalar value defined at the $i^{th}$ vertex. The interpolation is defined over all the vertices on the $k^{th}$ tetrahedron ($\mathcal{T}_k$) in which, the point $\mathbf{x}$ lies. Based on Eq.\eqref{eq:scalar_interpolation}, the gradient of $g(\mathbf{x})$ can be computed by
\begin{equation}\label{eq:gradient}
 \nabla g(\mathbf{x}) = \sum_{v_i \in \mathcal{T}_k \,:\,\mathbf{x} \in \mathcal{T}_k}  g_i \nabla h_i(\mathbf{x}).
\end{equation}
As $h_i(\mathbf{x})$ is a linear function, Eq.\eqref{eq:gradient} shows that the value of $\nabla g(\mathbf{x})$ is a constant inside each tetrahedron. This is well compliant with the representation of $\mathbf{v}(\mathbf{x})$ employed in our method.

\subsection{Vector Field Construction}
To make a vector field $\mathbf{v}(\mathbf{x})$ controllable, the user specified control vectors $\{\mathbf{a}_j\}_{j=1,2,\cdots,m}$ defined on the locations $\{\mathbf{x}_j\}_{j=1,2,\cdots,m}$ (with $\mathbf{x}_j \in \Omega$) need to be interpolated. The elements that contain these control vectors are stored in a set $\mathcal{C}(\Omega)$ (named as the constraint set). Then, the following term needs to be satisfied: 
\begin{equation}
    \sum_{\mathcal{T}_i \in \mathcal{C}(\Omega)} ||\mathbf{v}_i - \mathbf{a}_i||^2 = 0
    \label{eq:preserve}
\end{equation}
where $\mathbf{a}_i$ denote the target / anchor vector in the $i^{th}$ tetrahedron.

These anchor vectors are required to be smoothly interpolated by the vector field $\mathbf{v}(\mathbf{x})$. Therefore, the following objective function is defined to minimize the difference of the vectors in the neighbouring elements (assuming this local step can give rise to a similar global effect).
\begin{equation}
    \min_{\{ \mathbf{v}_i \}} \sum_{i=1}^{n_\mathcal{T}} \sum_{\mathcal{T}_j \in \mathcal{N}(\mathcal{T}_i)} \| \mathbf{v}_i-\mathbf{v}_j \|^2
    \label{eq:harmonic}
\end{equation}
where $\mathcal{N}(\mathcal{T}_i)$ represents the neighbour set of the $i^{th}$ tetrahedral element and $n_\mathcal{T}$ denotes the total number of tetrahedra in $\Omega$. This also introduces another important property -- the finally generated iso-surfaces from $g(\mathbf{x})$ are very smooth (by the virtue of an enforcing equality in the neighbouring surface normals). 

The requirements given in Eqs.\eqref{eq:preserve} and \eqref{eq:harmonic} are integrated to formulate an optimisation problem as:
\begin{equation}
    \min_{\mathbf{\{v_i\}}} \left( \alpha \sum_{i=1}^{n_\mathcal{T}} \sum_{\mathcal{T}_j \in \mathcal{N}(\mathcal{T}_i)} \|\mathbf{v}_i-\mathbf{v}_j\|^2 + \sum_{\mathcal{T}_i \in \mathcal{C}(\Omega)} \beta_i ||\mathbf{v}_i - \mathbf{a}_i||^2 \right)
    \label{eq:optimisation}
\end{equation}
where the terms $\alpha$ and $\beta_i$ are weights to be adjusted for setting the relative importance of each term. As there might have conflict between the target vectors as well as with the harmonic demand imposed in Eq.\eqref{eq:harmonic}, the weights are employed to decide the priority of different requirements. Detail discussion can be found in Sec.~\ref{subsubsecWeightTune}. 

This optimization problem can be solved by computing an over-determined linear equation system \cite{zhang2011robust} with $\mathbf{V}_{{n_\mathcal{T}} \times 3}$ being a matrix containing the three components of each vector on columns and the vectors on the rows. We use a similar matrix $\mathbf{V}_a$ to contain the anchor vectors. 
\begin{equation}
    \begin{bmatrix}
    \alpha \mathbf{L_u}\\
    \beta_i  \mathbf{I_a}
    \end{bmatrix}
    \mathbf{V}
     = \begin{bmatrix}
    0\\
    \beta_i  \mathbf{V}_a
    \end{bmatrix} 
    \label{eq:linear}
\end{equation}
where  $\mathbf{I_a}$ is an identity matrix and $\mathbf{L_u}$ is a Laplacian matrix as
\begin{equation}
    (L_u)_{i^{th} Row}\mathbf{V} = \mathbf{v}_i - \frac{1}{|\mathcal{N}(\mathcal{T}_i)|}  \sum_{\mathcal{T}_j \in \mathcal{N}(\mathcal{T}_i)} \mathbf{v}_j
\end{equation}
with the neighbour of the $i^{th}$ tetrahedron (denoted by $\mathcal{N}(\mathcal{T}_i)$) being the set of tetrahedra sharing a common face with $\mathcal{T}_i$. Although we present the three components of vectors in the same system in Eq.\eqref{eq:linear}, they are solved separately and normalized thereafter. 

  \subsection{Scalar Field Generation}
 Once the vector field $\mathbf{v}(\mathbf{x})$ has been determined in the domain $\Omega$, we can compute the scalar field $g(\mathbf{x})$ to satisfy $\nabla g(\mathbf{x}) = \mathbf{v}(\mathbf{x})$ ($\forall \mathbf{x} \in \Omega$). As aforementioned, the scalar field $g(\mathbf{x})$ can be determined by solving a Poisson's equation (Eq.\eqref{eq:v_div}) \cite{crane2013geodesics, liao2009gradient}.
 
 For a tetrahedral mesh, the Laplacian $\Delta$ can be computed as:
\begin{equation}
   \Delta   =  \mathbf{M}_{(n_\mathcal{V} \times n_\mathcal{V})}^{-1}  \mathbf{L}_{c{(n_\mathcal{V} \times n_\mathcal{V})}}
\end{equation}
where $\mathbf{M}$ is a diagonal matrix and $\mathbf{L}_c$ is the Co-tangent Laplacian for a tetrahedral mesh (see the Appendix for more details). Considering the discrete representation of the divergence of $\mathbf{v}(\mathbf{x})$ be represented as $\mathbf{D}_{(n_\mathcal{V} \times 1)}$, the Poisson's equation is solved (for $g(\mathbf{x})$) using the following linear system:
\begin{equation}\label{eq:discretePoisson}
\mathbf{L_c} \, g = \mathbf{M} \, \mathbf{D}.
\end{equation}
The discrete form of the vector operators (including gradient, divergence and Laplacian) on a tetrahedral mesh are also provided in the Appendix. The scalar field determined by Eq.\eqref{eq:discretePoisson} can vary by a constant, which can controlled by fixing the scalar values at selected points. We can impose Dirichlet boundary conditions on a surface (e.g., the convex hull of an input model) to indicate the `sources' of a field. The effectiveness of imposing such boundary conditions will be further studied in Sec.~\ref{subSecCompRes}.

\subsection{Surface extraction} \label{subSecSurfaceExtrct}
\label{surface extraction}
After obtaining the scalar field $g(\mathbf{x})$, we can extract the iso-surfaces of $g(\mathbf{x})$ to use as the working surfaces. As the normalized vectors are employed in our approach for the vector field $\mathbf{v}(\mathbf{x})$, the distances between iso-surfaces using constantly increased iso-values can be nearly uniform considering the errors between $\nabla g$ and $\mathbf{v}$ are trivial. 

The iso-surfaces are represented as a triangle mesh, which can be generated by locating all vertices with the specified iso-value $g_c$ on the edges of tetrahedral elements and forming triangles inside those elements with nodes having scalar values $>g_c$ and $\leq g_c$. Contour-parallel toolpaths, along which the cutting tool can traverse, are generated on each working surface (i.e., the triangular mesh of iso-surfaces) by using a geodesic field based method. Details can be found in \cite{fang2020reinforced}. 



\section{Vector Field Improvement}
\label{SecVFImprove}

Computing an optimized solution by Eq.~\eqref{eq:linear} can result in vector fields that approach the requirements stated in Proposition \ref{prop:curl} and \ref{prop:singularity}; however, there is no explicit term to penalize the formation of singular points or significant rotational components. Moreover, the second term in Eq.\eqref{eq:optimisation} introduces constraints on the vector field, where there is a chance to form a vector field significantly violating Proposition \ref{prop:curl} and  \ref{prop:singularity} if the anchor vectors are chosen inappropriately. Methods are presented in this section to remove singularities and rotational components from a vector field. 

\begin{figure}[t]
\centering 
\includegraphics[width=\linewidth]{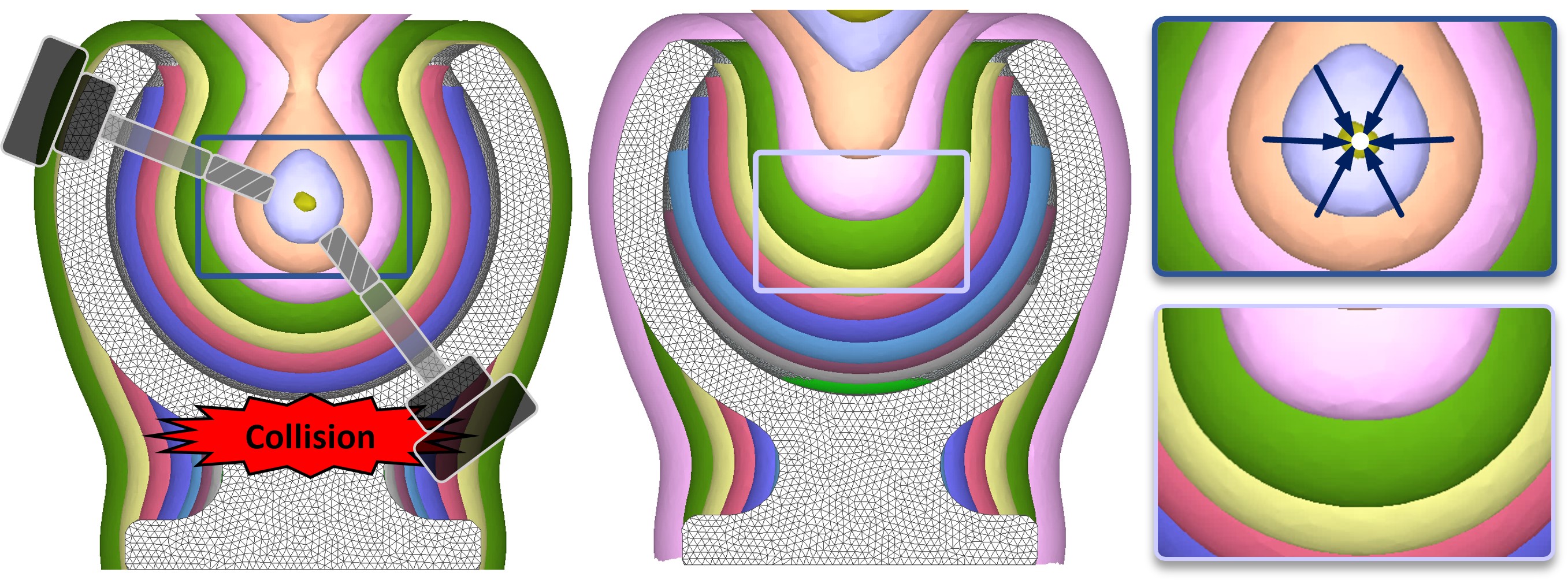}
\put(-223,-7){\small \color{black}(a)}
\put(-120,-7){\small \color{black}(b)}
\put(-39,-7){\small \color{black}(c)}
\caption{An illustration for removing Type I -- \textit{Point Singularity}. (a) Working surface layers formed with a singularity point in the center. Note that this singularity leads to floating volumes enclosed by the layers that cannot be machined collision-freely. (b) Processed peeling layers after removing the singular point with local maximal / minimal scalar value. (c) Comparison in zoom view for the layers before (top) vs. after (bottom) removing the point singularity.}
\label{fig:isolatedSingularity}
\end{figure}

\subsection{Singularity Processing}\label{subSecSingularityResol}
Four different types of singularity are considered below.

\subsubsection{Type I: Point Singularity}
In the scalar field $g(\mathbf{x})$ generated from the vector field $\mathbf{v}(\mathbf{x})$ by solving Poisson's equation, a local maxima (minima) can lead to a point singularity. Figure \ref{fig:isolatedSingularity}(a, c) illustrate a typical scenario of iso-surfaces around a point singularity. We name such singularity as \textit{Type I -- Point Singularity}. The floating volume generated around the singular point is not manufacturable, and it also leads to collision between the tool and the uncut material. 

To remove Type I singularity, we first identify every singular point by checking local maxima / minima in the scalar field $g(\mathbf{x})$. An anchor vector is introduced at the position of a singular point and points towards an accessible direction. In our current implementation, the direction is specified by users although it can also be assigned automatically by the heuristics suggested in \cite{spitz2000accessibility,chen2021design}. After adding new anchor points into the set of constraints, we need to generate the vector field again by solving Eq.~\eqref{eq:linear}. Singular points need to be detected and removed again on the newly generated vector field. The process is repeated till all the singular points have been removed. 

\subsubsection{Type II: Set Singularity -- Selected Surface}
\begin{figure}[t]
\centering 
\includegraphics[width=\linewidth]{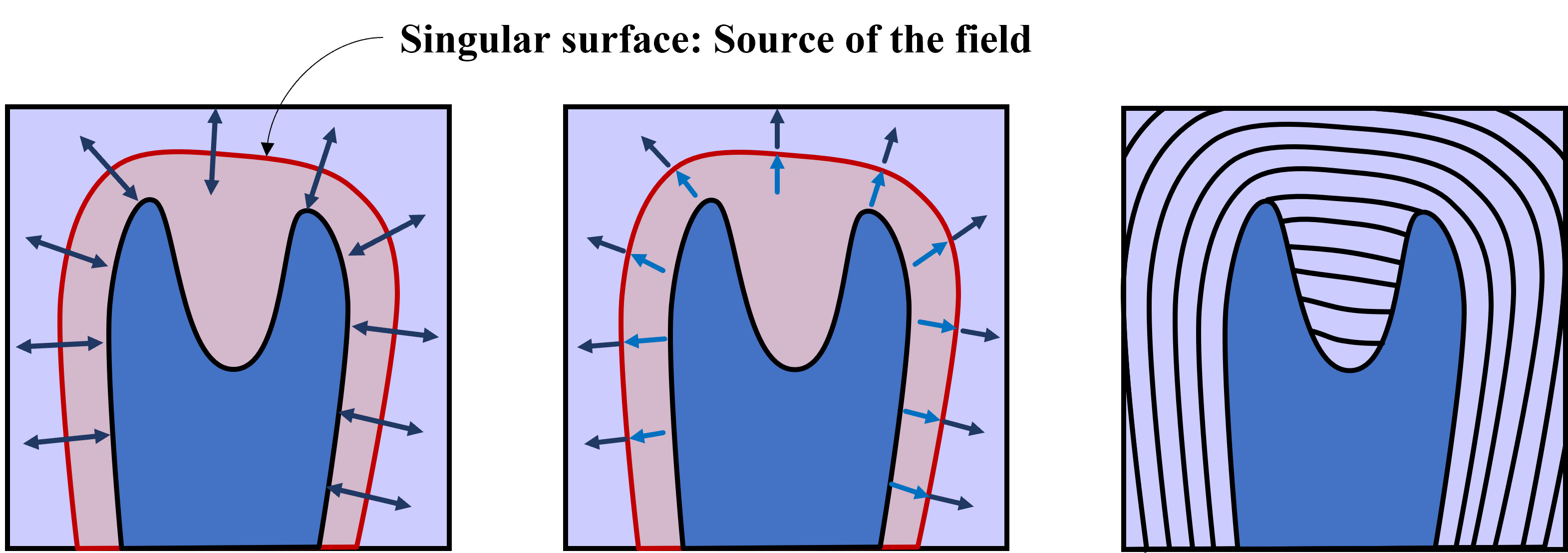}
\put(-225,-10){\small \color{black}(a)}
\put(-130,-10){\small \color{black}(b)}
\put(-40,-10){\small \color{black}(c)}
\caption{An illustration for Type II singularity and its removal. (a) The vector field initiating from a source surface, which is a convex hull surface as employed in \cite{he2021geodesic}. (b) The vectors on the interior side of the convex hull are flipped to make a continuous and compatible vector field for generating a scalar field without singularity. (c) Typical layers generated from the iso-surfaces on a scalar field with Type II singularity removed.}
\label{fig:Proposition_3_1}
\end{figure}

Local extremity in a scalar field can also appear in the form of a set of connected points as a surface (or a curve). An example of such a scenario is in the case of using the geodesic distance-field based method \cite{he2021geodesic}, where the convex hull of an input model is used as the sources (see Fig.\ref{fig:Proposition_3_1}(a)). Although such singularities (as sources) contain the information to generate a scalar field, the local maxima and minima formed by these singularities will form isolated volumes and also working surface layers with non-accessible points. 

When an oriented surface is given as a source, the problem of singularity can be easily solved by using only normals on one side of the surface to initiate the vector field. We refer to such a singularity as \textit{Type II}. This correction step expects the input surface to be `well-behaved' in properties like orientation and manifoldness. This has been depicted in Fig.~\ref{fig:Proposition_3_1}.

\begin{figure}[t]
\centering 
\includegraphics[width=\linewidth]{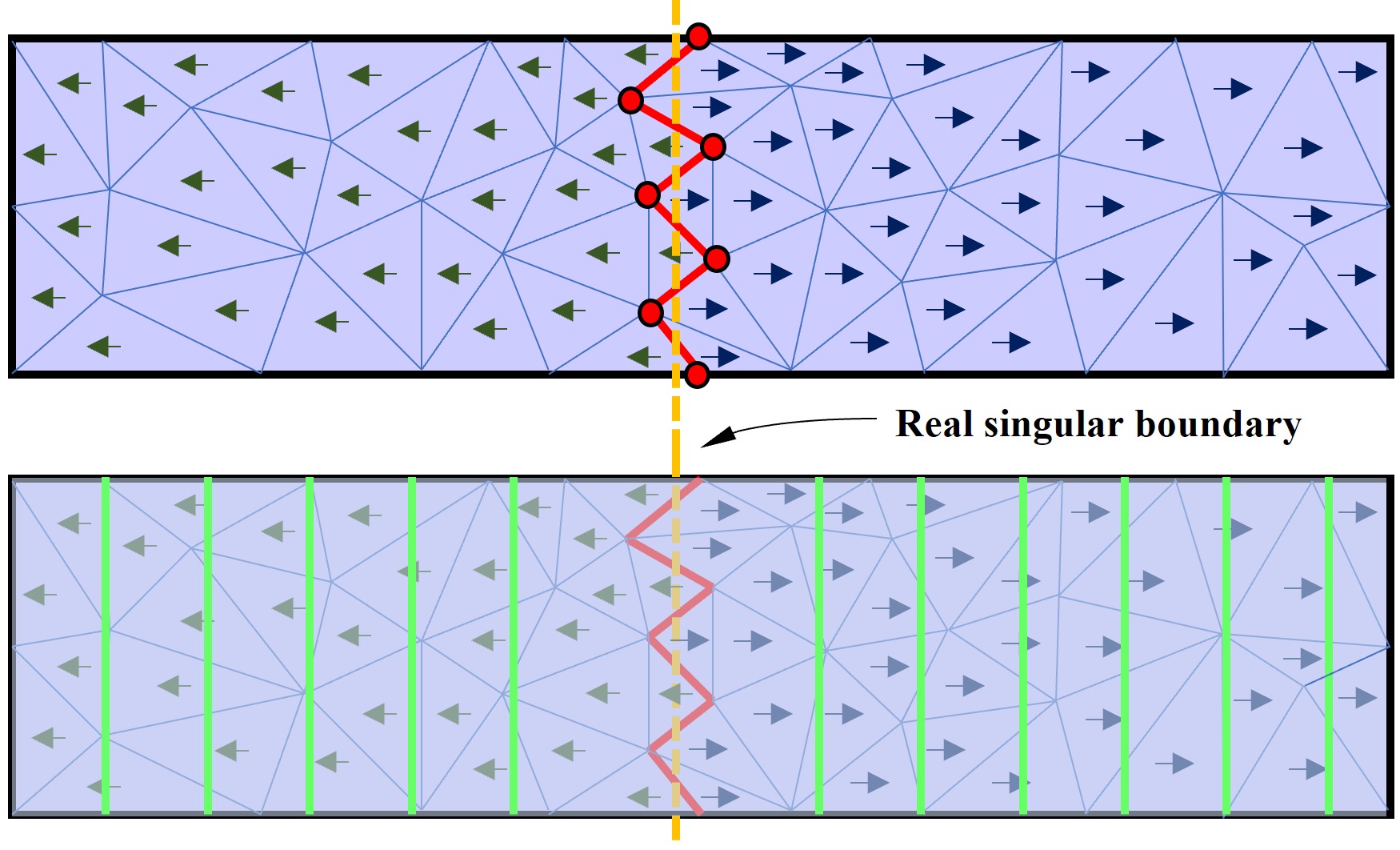}
\put(-252,154){\small \color{black}(a)}
\put(-252,73){\small \color{black}(b)}
\caption{(a) An example of planar singularity (Type III) in a meshed domain. The dashed line shows the real singular boundary whereas the red boundary is the discrete singular boundary across which the vectors point towards opposite directions. The red nodes define a boundary of in-compatibility as the vectors on either side demand the gradient to be in opposing directions which is conflicting. (b) The layers (green) are generated using the field in (a). The layers cannot be generated in the vicinity of the singularity zone due to in-compatible vectors.}
\label{fig:PlaneCorrection1}
\vspace{-10pt}
\end{figure}

\subsubsection{Type III: Set Singularity -- Admissible}\label{subsubsecType3Singularity}
Set singularity can also be generated by other reasons even if the points in the singular set are not selected as sources. Example cases are often found when there is conflict between multiple anchor vectors. In such cases, a new problem arise due to the discrete nature of our implementation. As highlighted in Fig.~\ref{fig:PlaneCorrection1}(a), the true singular surface and discrete singular surface are not the same. This gives rise to a discrepancy at the nodes adjacent to such a singular boundary as they lie on faces having conflicted gradients required in different sides. This should create a problem in computing a scalar field and extracting the layers around that region (see Fig.~\ref{fig:PlaneCorrection1}(b)), which is verified by our results discussed in Sec.~\ref{subsubSecResult_singular}).

\begin{figure}[t]
\centering 
\includegraphics[width=\linewidth]{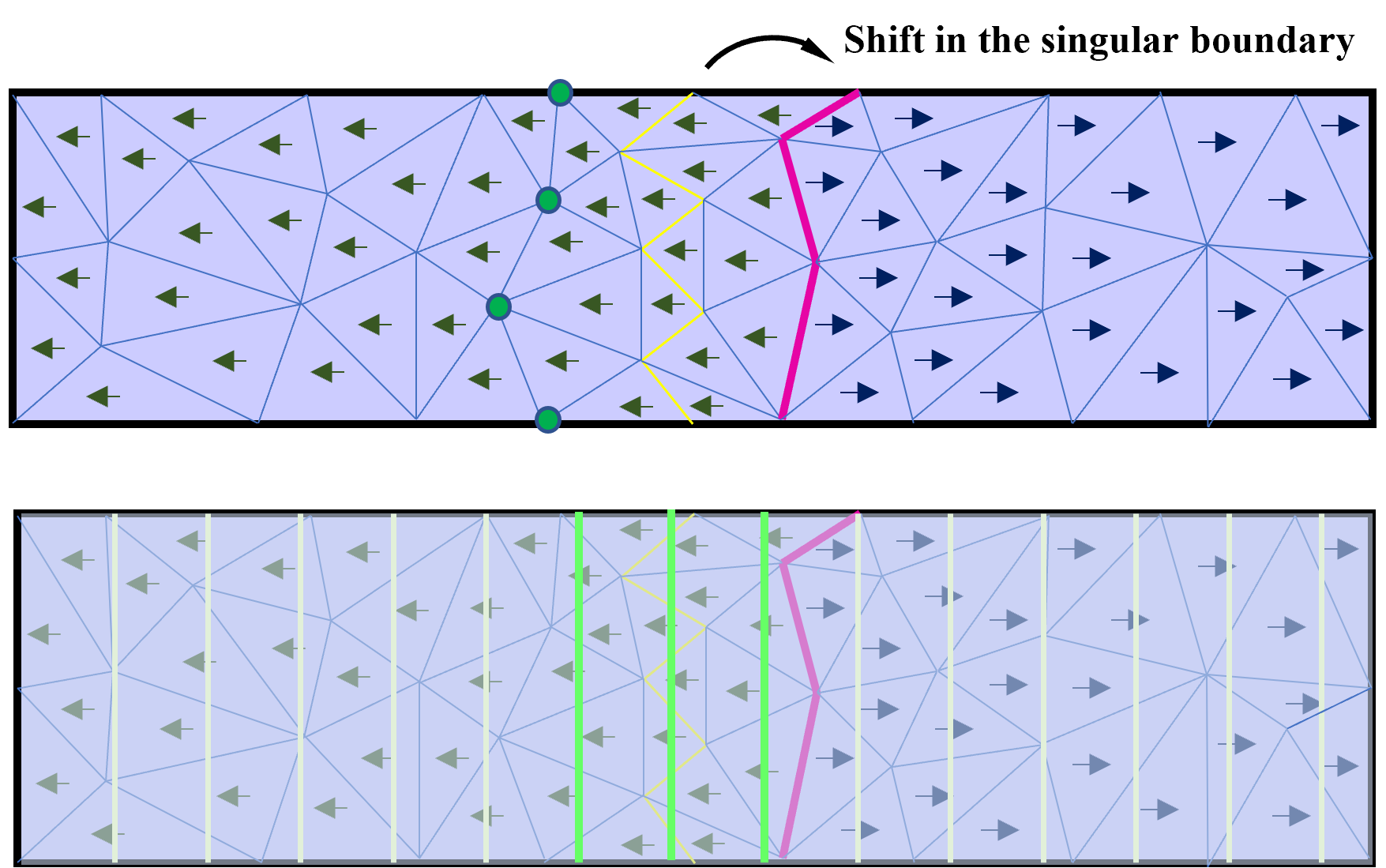}
\put(-252,150){\small \color{black}(a)}
\put(-252,70){\small \color{black}(b)}
\caption{The method for resolving \textit{Type III} singularity. (a) The original discrete singular boundary (yellow) is shifted to the new boundary (magenta) by flipping and smoothing the vectors on one side of the original boundary, where the green nodes on the other side are constrained to keep their original scalar values. (b) Layers (green) extracted at the original singularity zone. These layers are then added to the set of layers extracted from the original state of a vector field.}
\label{fig:PlaneCorrection2}
\end{figure}

We develop a \textit{local correction} method to process this type of singularity. The region with the occurrence of such a discrepancy is first detected by checking the angle between the vectors across each face -- all the faces with discrepancy form a \textit{singular boundary}. When conflict occurs, we flip and smooth the vectors in the neighboring region of the singular boundary. This, therefore, pushes the singular region towards one direction (e.g., towards the right in Fig.~\ref{fig:PlaneCorrection2}(a)). On the other side of the singular boundary, the scalar values of the nodes are constrained to their previously computed values (e.g., the green nodes displayed in Fig.~\ref{fig:PlaneCorrection2}(a)). Then a new scalar field is fitted and the layers corresponding to the iso-scalar values that were in the singularity region in the original field are extracted (see Fig.~\ref{fig:PlaneCorrection2}(b)). This is repeated once again by pushing the singularity in the other direction and extracting the same scalar valued layers again. Finally, the layers affected due to singularity in the original set of extracted surfaces are replaced by the new ones. The singularities which can be resolved using this method will be referred to as \textit{Type III}. As will discuss in the following paragraph, this type of singularity should satisfy a few conditions on their boundary.


\begin{figure}[t]
\centering 
\includegraphics[width=\linewidth]{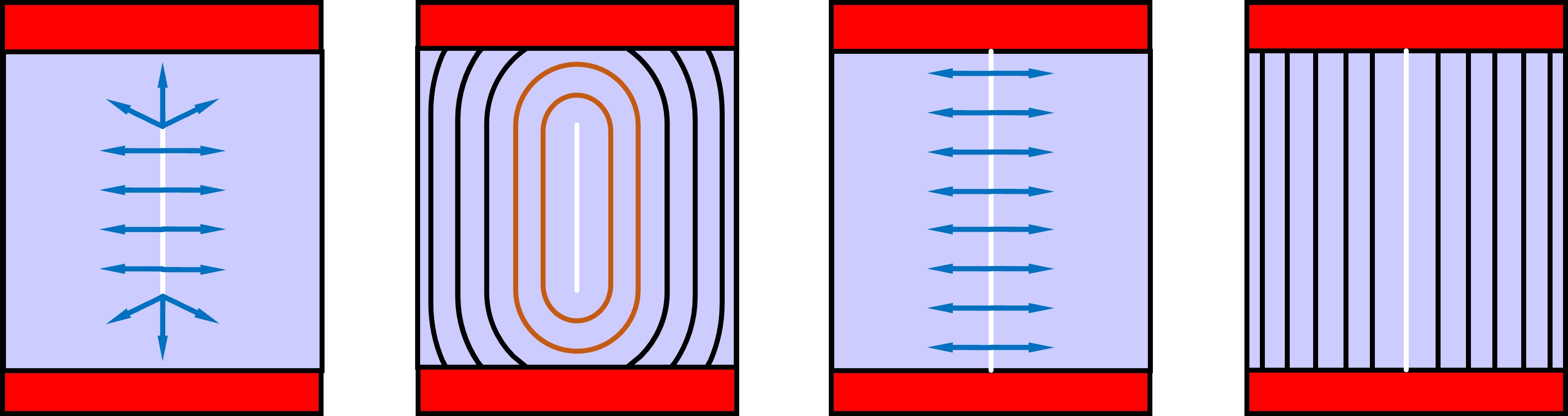}
\put(-235,-8){\small \color{black}(a)}
\put(-220,10){\small \color{black}$\Omega$}
\put(-220,60){\small \color{white}$\mathcal{H}$}
\put(-165,-8){\small \color{black}(b)}
\put(-100,-8){\small \color{black}(c)}
\put(-33,-8){\small \color{black}(d)}
\caption{
A comparison to show the difference between the \textit{Type IV} singularity (a) and its corresponding iso-surfaces (b) vs. the \textit{Type III} singularity (c) and iso-surfaces (d). Only when the surface singularity's end overlaps with the boundary of $\Omega$ and $\mathcal{H}$ as (c), the singularity can be well processed by the method in Sec.\ref{subsubsecType3Singularity}.
}\label{fig:Proposition_3}
\end{figure}

\begin{figure}[t]
\centering 
\includegraphics[width=\linewidth]{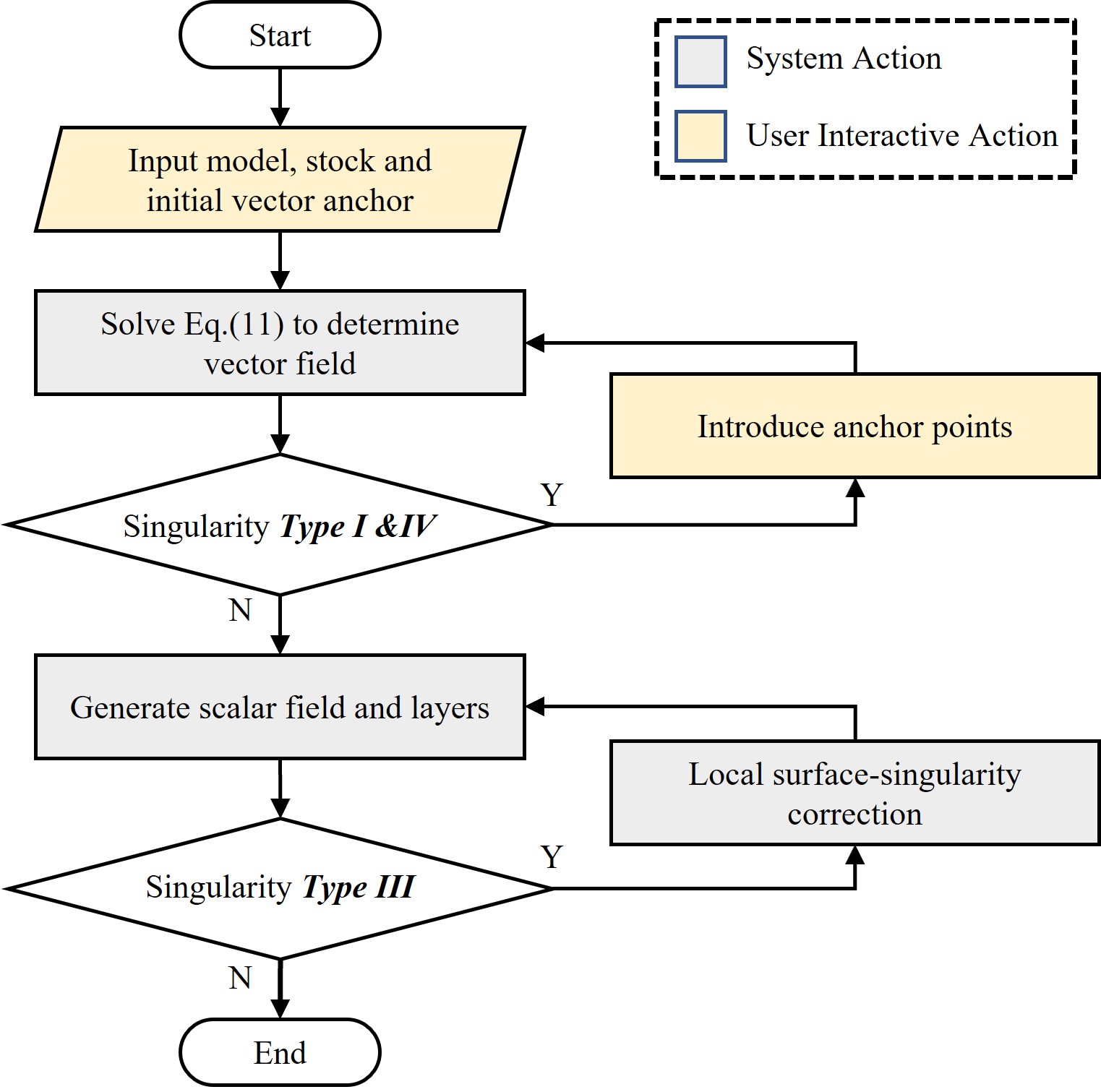}
\caption{The diagram of our singularity processing algorithm.}
\label{fig:algorithm_singularRmv}
\vspace{-10pt}
\end{figure}

\subsubsection{Type IV: Set Singularity -- Non-Admissible}
There is another type of surface singularity that cannot be repaired by the above method. As shown in Fig.~\ref{fig:Proposition_3}(a, b), when the singular manifold  has its ends (or generally speaking its boundary) inside the domain $\Omega$, the floating volume will be formed. Differently, for the Type III singularity as shown in Fig.~\ref{fig:Proposition_3}(c), the nature of layers generated as depicted in Fig.~\ref{fig:Proposition_3}(d) satisfy our definition of manufacturability. This condition can be generalized as the requirement that such singularities should have their entire boundary overlapping with the common boundary of the to-be-removed volume  $\Omega$ and the model domain $\mathcal{H}$. If not, the singularity needs to be either forced to satisfy that condition or removed, both of which can be realized by adding more anchor vectors. This class of singularities is referred to as \textit{Type IV}.

\vspace{8pt} Though \textbf{Proposition} \ref{prop:singularity} demanded the vector field to be completely singularity free, we have shown that some singularities can be admitted without deviating from the manufacturability requirements, thereby making it less restrictive. In fact, the type of singularities we desire to admit still does not violate the proposition, as such singularities can be thought of as the location of the imaginary boundary of the domain $\Omega$. This makes the proposition still valid inside the domain. Due to the large variation of the occurrence and admissibility of the singularities, the processing becomes a complex task. At this point, we aim to benefit from an interaction with the user to resolve the problems. The diagram of our algorithm is presented in Fig.\ref{fig:algorithm_singularRmv}.

         

\subsection{Curl removal} \label{subSecCurlRemov}
While the output of Eq.\eqref{eq:linear} gives an optimized vector field according to the objectives discussed in Eqs.\eqref{eq:preserve} and \eqref{eq:harmonic}, there is no guarantee that the resultant field satisfies the criterion of irrotational as stated in \textbf{Proposition} \ref{prop:curl} -- see Fig.\ref{fig:ProblemFormulation} for an example. This often happens when adding a new anchor vector that is too close to another anchor vector with a large variation in direction. Our observation finds that a satisfactory shape of isosurfaces can be obtained when the value of metric $I_{rot}$ is less than $4 \times 10^{-4}$.  The term $I_{rot}$ can be thought of as the mean-square of the error $(\mathbf{v}-\nabla g)$. When this is not satisfied, an iterative algorithm is employed to reduce the  rotational component of the designed vector field.

Instead of directly minimizing the value of $\| \nabla \times \mathbf{r} \|$, we employ a more computation-effective way. Specifically, after determining a scalar field $\phi(\mathbf{x})$ by solving the Poisson's equation $\nabla \cdot \mathbf{v} = \Delta \phi$ and natural boundary condition, we can evaluate the value of $\| \nabla \times \mathbf{r} \|$ by $\| \mathbf{v} - \nabla \phi \|$ according to Eq.\eqref{eq:HodgeDecomp2}. Therefore, our algorithm iteratively minimizes the difference between $\mathbf{v}$ and $\nabla \phi$ by the following steps.
\begin{itemize}
\item Step 1:  Consider $\phi$ as the previous estimated scalar field.  

\item Step 2: We define a new energy term, by adding $\sum_{\mathcal{T}_k \in \mathcal{T}}\|\mathbf{v_k}-\frac{\nabla \phi_k}{\|\nabla \phi_k\|}\|^2$ to the one in Eq.\eqref{eq:optimisation} as:
\begin{equation}
  \begin{aligned}
  & \alpha \sum_{i=1}^{n_\mathcal{T}} ||\sum_{\mathcal{T}_j \in \mathcal{N}(\mathcal{T}_i)} (\mathbf{v}_i-\mathbf{v}_j)||^2 + \sum_{\mathcal{T}_i \in \mathcal{C}(\Omega)} \beta_i ||\mathbf{v}_i - \mathbf{a}_i||^2 
  \\ & + \gamma  \sum_{\mathcal{T}_k \in \mathcal{T}}\|\mathbf{v_k}-\frac{\nabla \phi_k}{\|\nabla \phi_k\|}\|^2
  \label{eqn:optimiseCurl}
  \end{aligned}
\end{equation}

\item Step 3: Consider the critical anchor vectors ($\mathbf{a_k}$) as the ones defined by the user to guide accessibility and remove singularity, we update the constraint set, $\mathcal{C}(\Omega)$, as set of those anchors. 

\item Step 4: Highest weight $\beta=10^8$ is allocated to those associated with critical anchors, followed by $\gamma = 10^5$ and then $\alpha=1.0$. More discussion about the weights can be found in Sec.\ref{subSecDiscussion}.

\item Step 5: Minimise the new energy and update scalar field $\phi$.

\item Step 6: Compute the value of metric $I_{rot}$ and go back to Step 1 if the error is greater than $4 \times 10^{-4}$.
\end{itemize}
When the computation converges, the vector field $\textbf{v}(\textbf{x})$ has been converted into an irrotational one $\Tilde{\textbf{v}}(\textbf{x})$.

The approach presented above is based on the fact that the direction of the gradient of the scalar field obtained after first iteration, gives us a heuristic about the vector distribution that can be estimated well by a gradient field. Using these directions should naturally relax the field that are leading to a sharp turn. In our approach, we only enforce the magnitude constraint on those directions. However, if any other constraint must be applied, it can be reintroduced into the new vector field as anchors and be interpolated while preserving the original distribution using the identity part of Eq.\eqref{eq:linear}.




\section{Results and Discussion}

\subsection{Implementation details} \label{subsecImplementation}
The process is implemented in the form of an interactive tool. The code is written in C++ along with the use of openGL and Qt to build an interactive interface. The program is run on a PC with 3.80GHz Intel Core i7 (16 Logical core) processor with 32.0 GB of RAM. The PC has a NVIDIA GeForce 3070 (8.00 GB dedicated memory) GPU. We use the Eigen library\cite{guennebaud2010eigen} for the vector operations and solutions which is accelerated using Intel MKL \cite{wang2014intel}. The PQP library \cite{gottschalk1996obbtree} is used when the signed distances from query points to the given surface model need to be determined. The supplementary video of our work can be accessed at: \url{https://youtu.be/Bzt2oe6YYh8}

\begin{figure}[t]
 \centering 
\includegraphics[width=\linewidth]{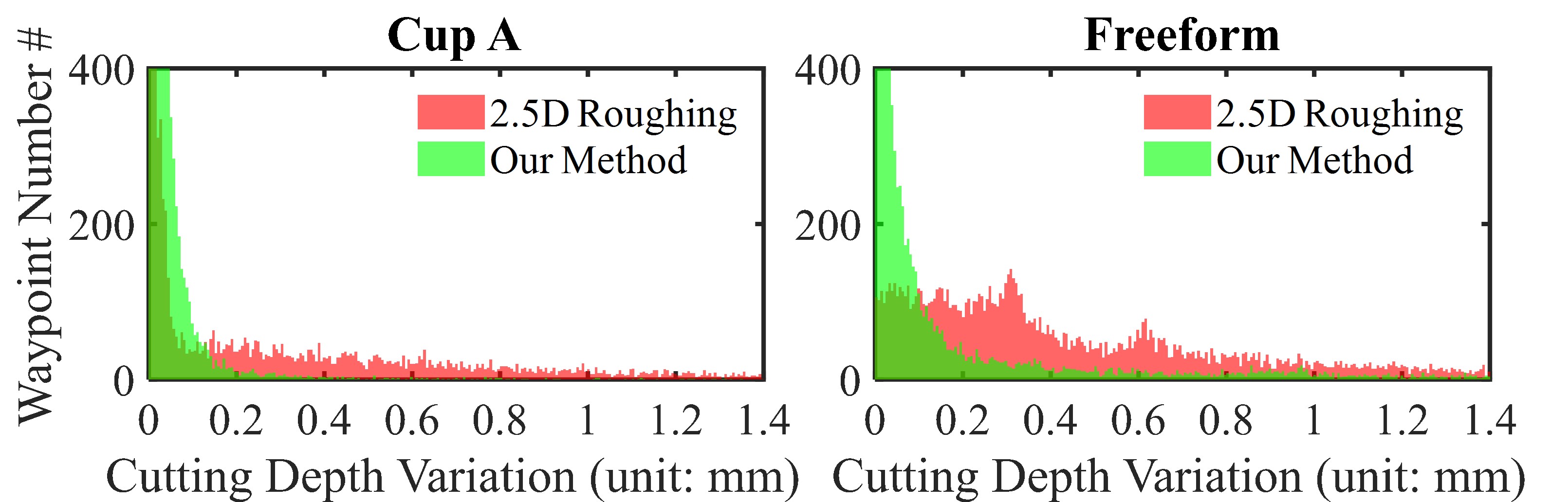}
\\
\small
\vspace{4pt}
\resizebox{\linewidth}{!}{\small
\begin{tabular}{r||c|c||c|c }
\hline
 (Unit: mm) & \multicolumn{2}{c||}{Maximal Cutting Depth} & \multicolumn{2}{c}{Average Cutting Depth Variation}\\
\hline 
Models & 2.5D Roughing & Our Method & 2.5D Roughing & Our Method \\
\hline 
\hline
Mannequin & 3.53 & 0.64 & 1.95 & 0.19\\
Cup A & 5.23 & 0.34 & 2.97 & 0.05\\
Freeform & 2.89 & 0.54 & 2.01 & 0.21\\
\hline 
\end{tabular}
}
\caption{The variation of cutting depth during finishing (by contour parallel toolpaths) on the 2.5D roughing results (planar peeling) vs. our roughing results (curved peeling), where abrupt change in cutting depth leads to negative dynamic effects on the process. The comparison is taken on both the histogram and the maximal / average values. 
}\label{fig:resCmp_cup}
\end{figure}

\begin{figure}[t]
\centering 
\includegraphics[width=\linewidth]{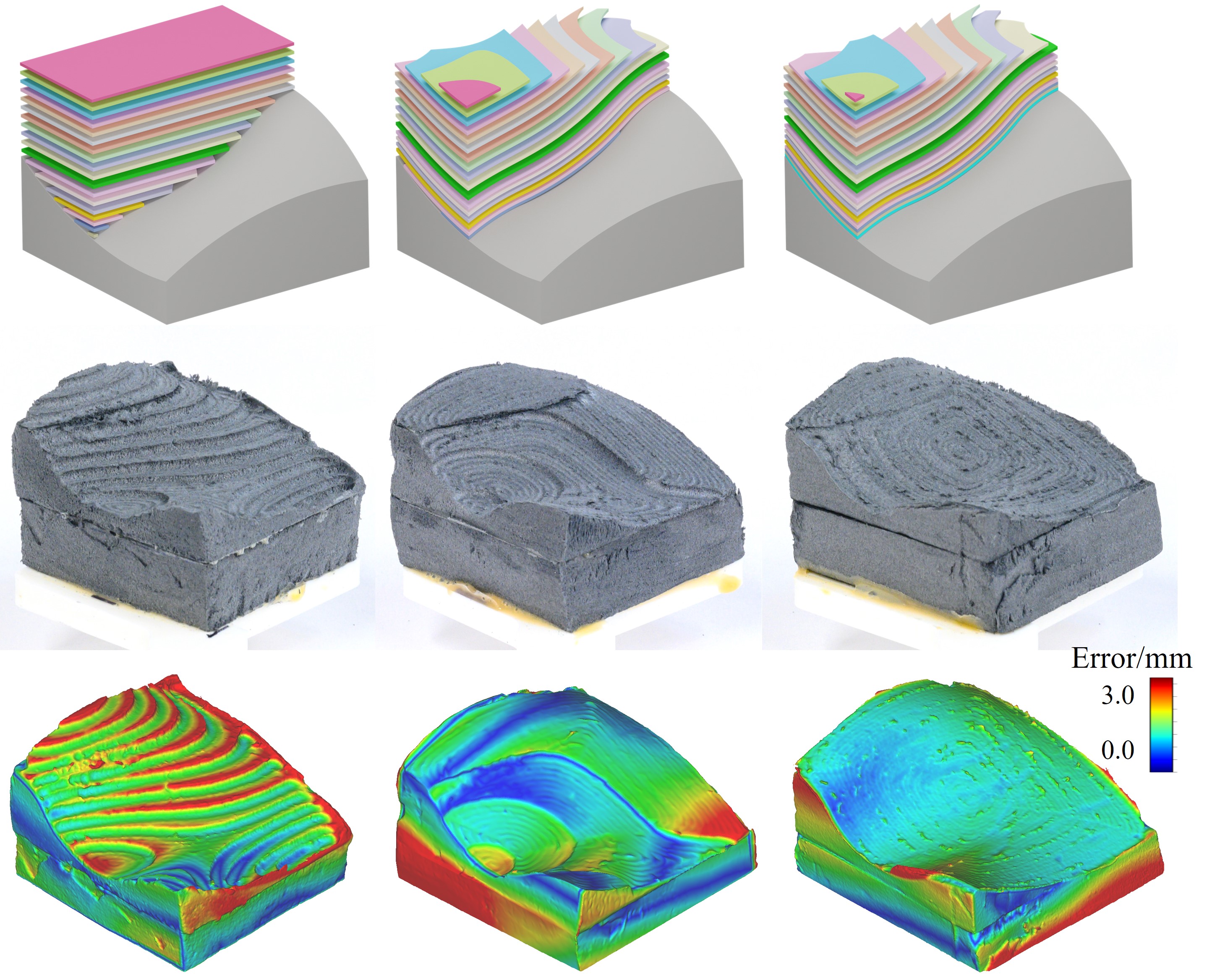}
\put(-257,139){\small \color{black}(a1)}
\put(-172,139){\small \color{black}(b1)}
\put(-87,139){\small \color{black}(c1)}
\put(-257,72){\small \color{black}(a2)}
\put(-172,72){\small \color{black}(b2)}
\put(-87,72){\small \color{black}(c2)}
\put(-257,2){\small \color{black}(a3)}
\put(-172,2){\small \color{black}(b3)}
\put(-87,2){\small \color{black}(c3)}
\caption{The comparison of the layers and results generated by using different roughing strategies -- (a) 2.5D roughing, (b) our curved peeling method with the field generated from natural boundary conditions and (c) our method by field generated from Dirichlet boundary condition. (Top row) layers for roughing, (middle row) results of physical fabrication, and (bottom row) the shape approximation errors obtained by scanning the physical results and comparing with the target model. Note that results of all the three strategies are produced by our framework but using different anchor vectors and boundary conditions.
}\label{fig:resFreeform}
\end{figure}
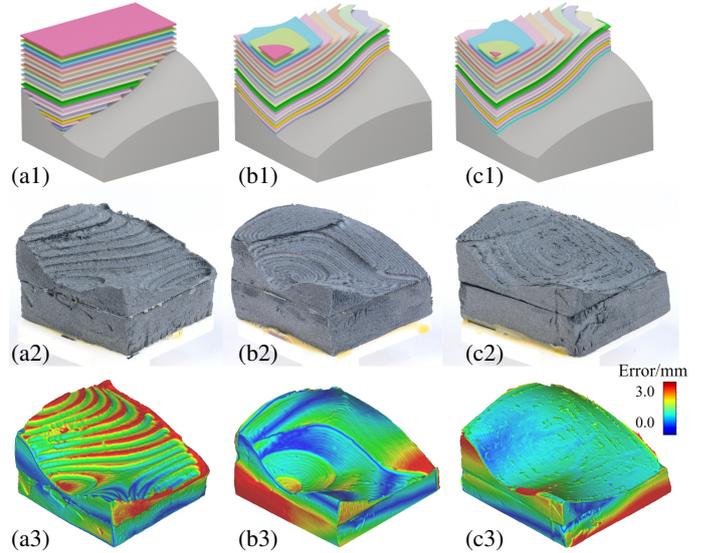

The vector design process requires an initial vector to be defined at at-least one point on the domain. This process is benefited by allowing the user to interactively select anchor points (constraints) and anchor vectors, and then change the peeled layers. In our case, a vector direction is meant to represent the desired tool orientation at that point. We allow the user to select a vertex in the domain $\Omega$ and then use the mouse to draw a vector direction or through keyboard input. The vector introduced on a vertex, translates to the target vector directions (which are added to $\mathbf{V}_a$ in Eq.(\ref{eq:linear})) on all the tetrahedral elements incident to that vertex. A typical process of controlling the layer shape by adding more constraint points using vertices is shown in Fig.~\ref{fig:pipeline}(b). Another way to assign the vectors is through the faces. Here the user is given the option to select the faces (specially the boundary faces of the model and the stock material) and then transfer the constraint into the tetrahedral element containing that face. Still another way of using faces is when a surface is introduced by the user (for instance, the convex hull) to initialize the vector field. In that case, the intersection tetrahedrons are computed and the surface normals of the intersecting faces elements of the other surfaces are transferred into the tetrahedrons (in accordance with the singularity resolution method). 

In our current version, for all the constraints provided through a surface (faces), we use the normal of these surfaces (faces) as the initialized vectors. This makes the peeling result conformal to these surfaces, which reduces the stair-case effect at these regions. All the anchor vectors assigned to either nodes or faces are transferred to the vector field of the associated tetrahedral elements. The final result of our optimization for curl removal is a globally optimized vector field by incorporating all such local constraints. The exact behaviour of the resultant vector field is thus influenced by how and where these constraints are imposed. Hence, the number of such constraints varies based on the tasks and requirement specified by users. As highlighted in Table~\ref{tab:comprarison}, the numbers of constrained elements are different in different cases. In practice, a user can always start from a few constraints / anchors and then progressively add more in the regions that do not provide the desired shape of peeling. 


\subsection{Computational results} \label{subSecCompRes}
The entire pipeline is represented in Fig.~\ref{fig:pipeline}. The calculation uses a tetrahedral mesh of the volume $\Omega$ ($\mathcal{M}-\mathcal{H}$ in Fig.~\ref{fig:pipeline}(a)). A field of vectors in $\Omega$ (Fig.~\ref{fig:pipeline}(c)) is used to represent the normal direction of the layers at each point. Assuming, the vector field satisfies the requirements set as a proposition, a scalar field in Fig.~\ref{fig:pipeline}(d) is computed (using Eq.\eqref{eq:discretePoisson}) such that the gradient of the scalar field is represented by the vector field. Finally, iso-surfaces (Fig.~\ref{fig:pipeline}(e)) of the scalar field are extracted, on which a cutting tool is  traversed to remove the material between two such surfaces. The process of obtaining the final vector field is shown in Fig.~\ref{fig:pipeline}(b). The left part of Fig.~\ref{fig:pipeline}(b) shows a set of initial anchor vectors (in this case, originate from the convex hull). After solving Eq.\eqref{eq:linear} gives a complete vector field in $\Omega$(the center part of Fig.~\ref{fig:pipeline}(b)). This vector field does not necessarily contain all the desired properties. For instance, the associated figure with the corresponding layers shows that near the base of the cup, the layers point downwards. This will prevent strong accessibility when the stock material is placed on a platform. Also, the layers in the cavity of the cup are flat, which will also inhibit accessibility. To solve this, we illustrate the use of our interactive interface, which allows the user to add local constraints in form of anchor vectors or by modifying the original field (the right part of Fig.\ref{fig:pipeline}(b)). This is done as many times as required by the user to obtain the desired layers. 

\begin{figure*}
    \centering
    \includegraphics[width = \linewidth]{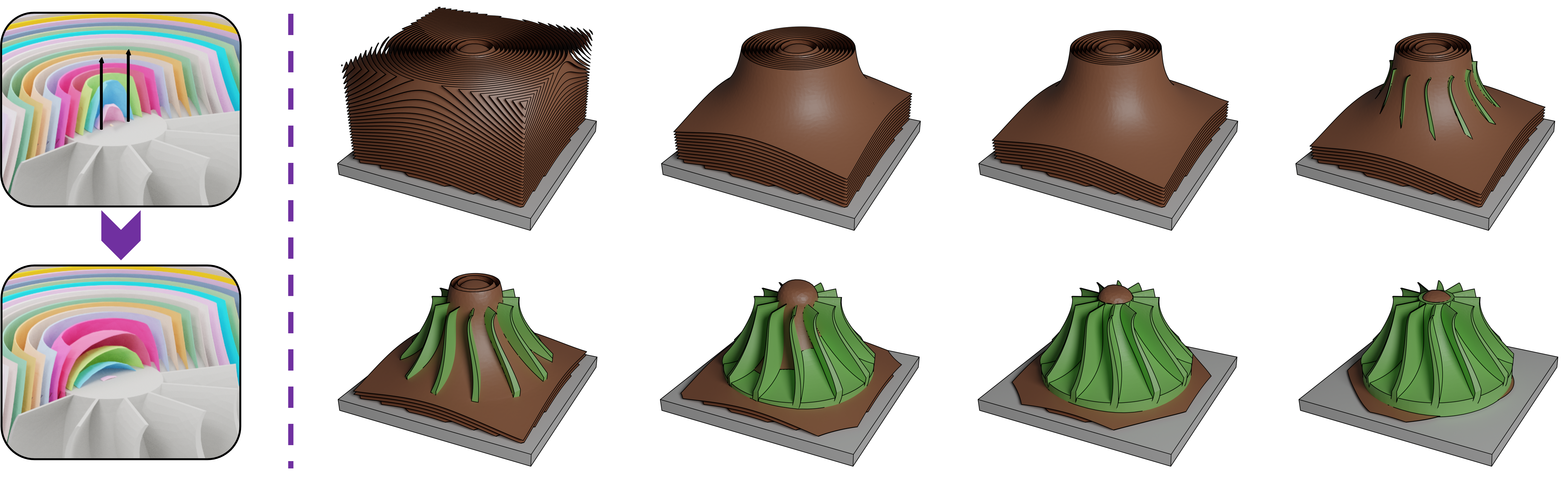}
    \put(-525,-1){\small \color{black}(a2)}
    \put(-525, 85){\small \color{black}(a1)}
    \put(-405,-1){\small \color{black}(b5)}
    \put(-405,85){\small \color{black}(b1)}
    \put(-295,-1){\small \color{black}(b6)}
    \put(-295,85){\small \color{black}(b2)}
    \put(-185,-1){\small \color{black}(b7)}
    \put(-185,85){\small \color{black}(b3)}
    \put(-75,-1){\small \color{black}(b8)}
    \put(-75,85){\small \color{black}(b4)}
\caption{The peeling results generated by our approach for an impeller model. For this example, the vector field was first generated by imposing anchor vectors according to the surface normals of the rotor body. After that, two more anchors -- the black arrows shown in (a1) are inserted at the top of the rotor to reduce the curvature of the layers on the top (a2). The layers of volume peeling are shown in (b1-b8) to demonstrate the progressive results to obtain the final model from a piece of stock material in cuboid-shape.}
    \label{fig:resCmp_impeller}
    \vspace{10pt}
\end{figure*}
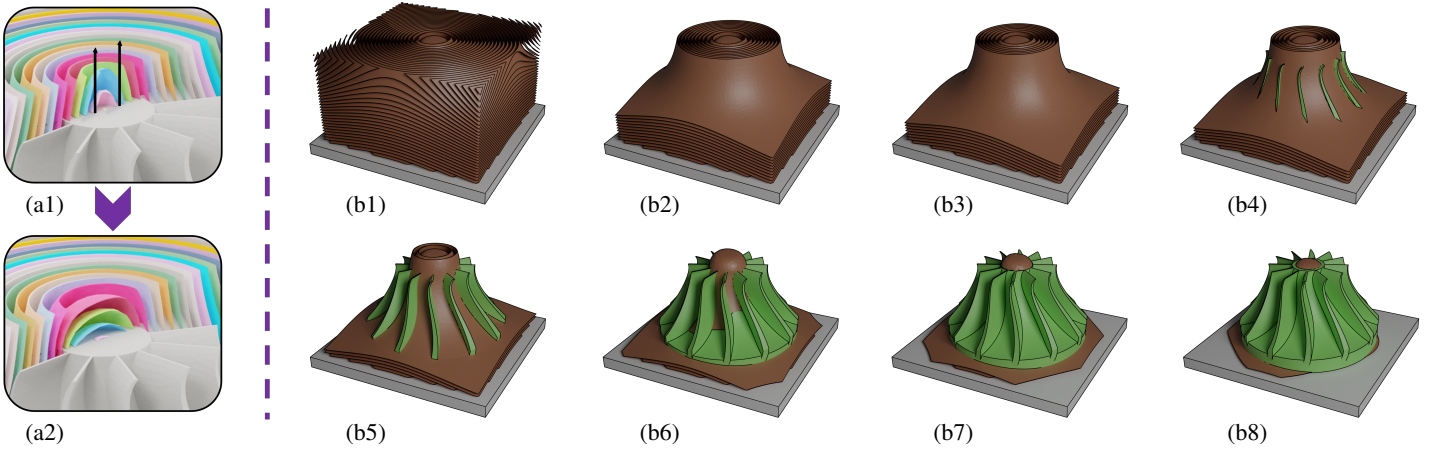

Figure~\ref{fig:resFreeform} shows the comparison of results generated by using different roughing strategies for a freeform surface. In Fig.~\ref{fig:resFreeform}(a1), a planar peeling is conducted to obtain the traditional 2.5D layers. In Figs.\ref{fig:resFreeform}(b1) and (c1), the vector fields are generated by our method but using different boundary conditions when solving the Poisson's equation to compute the scalar field from the vector field. As we know the final target surface in this case, we use a Dirichelt boundary condition in Fig.~\ref{fig:resFreeform}(c) to force the last layer to coincide with the final target surface. Differently, the natural boundary condition is employed for computing the result shown in Fig.~\ref{fig:resFreeform}(b). As a popular example for five-axis milling, we tested our approach on an impeller model as shown in Fig.~\ref{fig:resCmp_impeller}. It can be found that our method works well on models with complex geometry.

The results of the physical tests for these cases will be discussed in Sec.~\ref{subsecPhysicalExp}. 
The special cases relating to the different aspects of the vector field properties are discussed in the following sub-sections (i.e., Sec.~\ref{subsubSecResult_singular} and Sec.~\ref{subsubSecRes_curl}). Also, statistics of our method on a variety of models are given in Table~\ref{tab:comprarison}.

\begin{table}[t]
\small
\caption{Statistics of Computational Results}
\vspace{2pt}
\resizebox{\linewidth}{!}{\small
\begin{tabular}{r||c|c||c|c }
\hline
    Model & Total & Constrained  & Vector interpolat. & Poisson solution \\ 
     & tets\# & tets\# & time (sec.) &  time (sec.) \\
    \hline
    Cup A  & 424k & 31k & 40.8& 3.6\\
    Torus  & 391k & 10 & 27.2 & 3.1\\
    Mannequin  & 433k & 38k & 34.1& 3.3\\
    Freeform$^\dag$   & 390k & 3.9k & 56.4 & 3.5  \\
\hline
\end{tabular}
}
\begin{flushleft}
$^\dag$~The data of the Freeform model is calculated under the Dirichlet boundary condition.
\end{flushleft}
\label{tab:comprarison}
\vspace{-22pt}
\end{table}

\subsubsection{Singularity} \label{subsubSecResult_singular}
Figure \ref{fig:isolatedSingularity}(a) shows a case for the cup model, when the field is initialized from the surface normals of the model. Since, inside the cavity, the normals are directed toward the center, a singularity is formed in the center as depicted by the closed layers formed around it. We use the method described in Sec.~\ref{subSecSingularityResol} to remove the \textit{Type I} and \textit{Type IV} singularities by adding anchor vectors and obtain a singularity-free peeling as shown in Fig.\ref{fig:isolatedSingularity}(b).

\begin{figure}[t]
\centering 
\includegraphics[width=\linewidth]{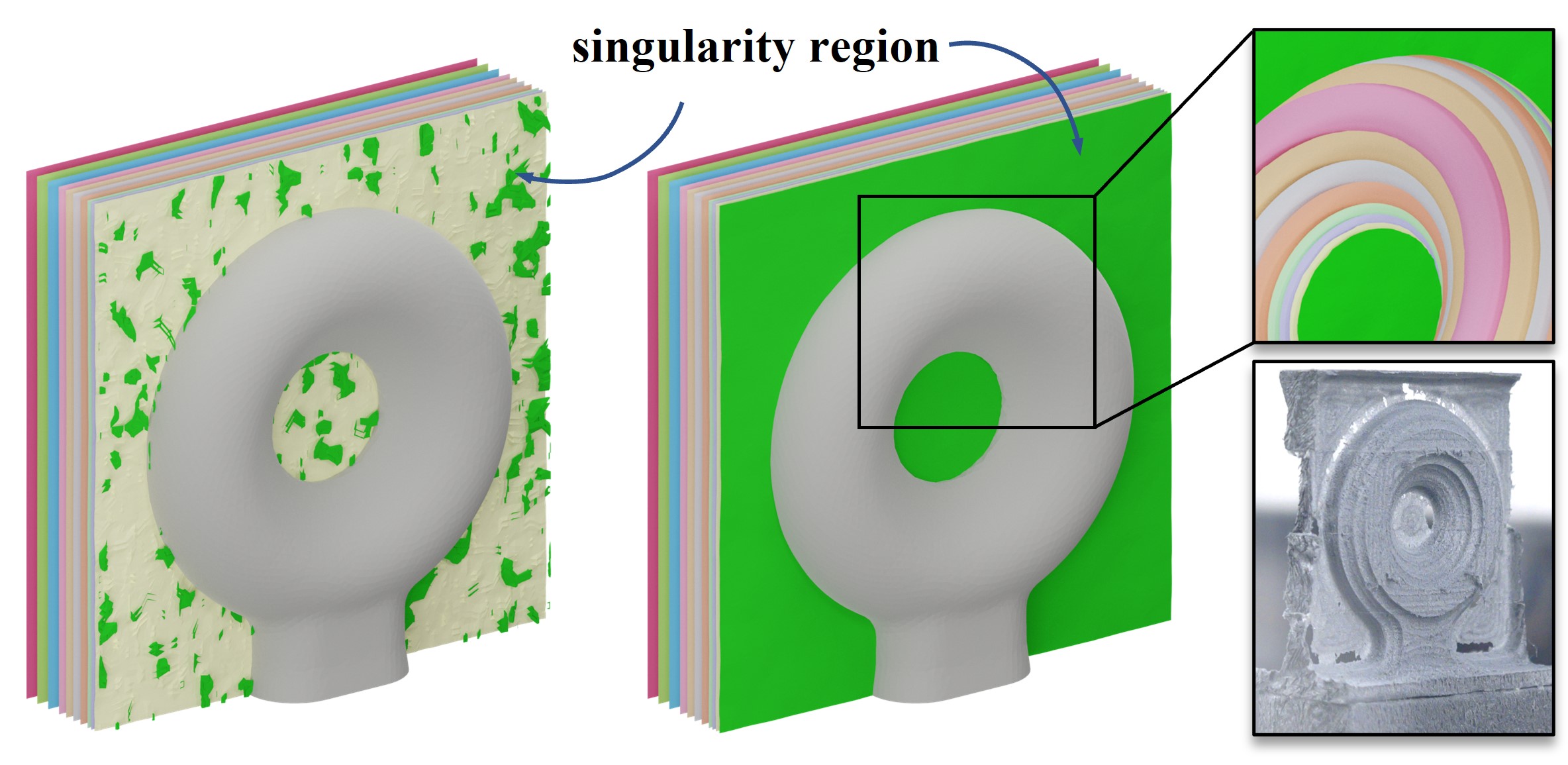}
\put(-254,-1){\small \color{black}(a)}
\put(-162,-1){\small \color{black}(b)}
\put(-65,-1){\small \color{black}(c)}
\caption{Processing layers in the singularity region, where (a) is before resolving incompatible layers due to local singularity and (b) is after running the local-correction algorithm presented in Sec.\ref{subSecSingularityResol}. (c) Zoom view on the figure of inner layers and the corresponding photo taken while machining the layer in the singular region.}
\vspace{-10pt}
\label{fig:resCmp_dryer}
\end{figure}

Another case of the singularities is that of the \textit{Type II} in which a selected surface is given as a source. Figure~\ref{fig:pipeline}(b) shows that by using only one side of the surface, such singularities can be avoided.  An example of the correction method depicted in Fig.~\ref{fig:PlaneCorrection2} for \textit{Type III} singularities is shown for a torus model in Fig.~\ref{fig:resCmp_dryer}. Figure~\ref{fig:resCmp_dryer}(a) shows the broken layers in the singularity zone which are corrected to obtain the layers shown in Fig.~\ref{fig:resCmp_dryer}(b). 
As illustrated in Fig.~\ref{fig:PlaneCorrection1}, this is because the elements of the discrete model in general do not align along the real singular surface. This adds noises / errors in the resultant scalar field, which leads to the broken patches in Fig.~\ref{fig:resCmp_dryer}(a). This problem on fields is corrected by the method explained in Fig.\ref{fig:PlaneCorrection2}. 
A zoom-in figure of the machining layer at the singular region after correction is shown in Fig.~\ref{fig:resCmp_dryer}(c).

\subsubsection{Curl removal} 
\label{subsubSecRes_curl}
\begin{figure}[t]
\centering 
\includegraphics[width=\linewidth]{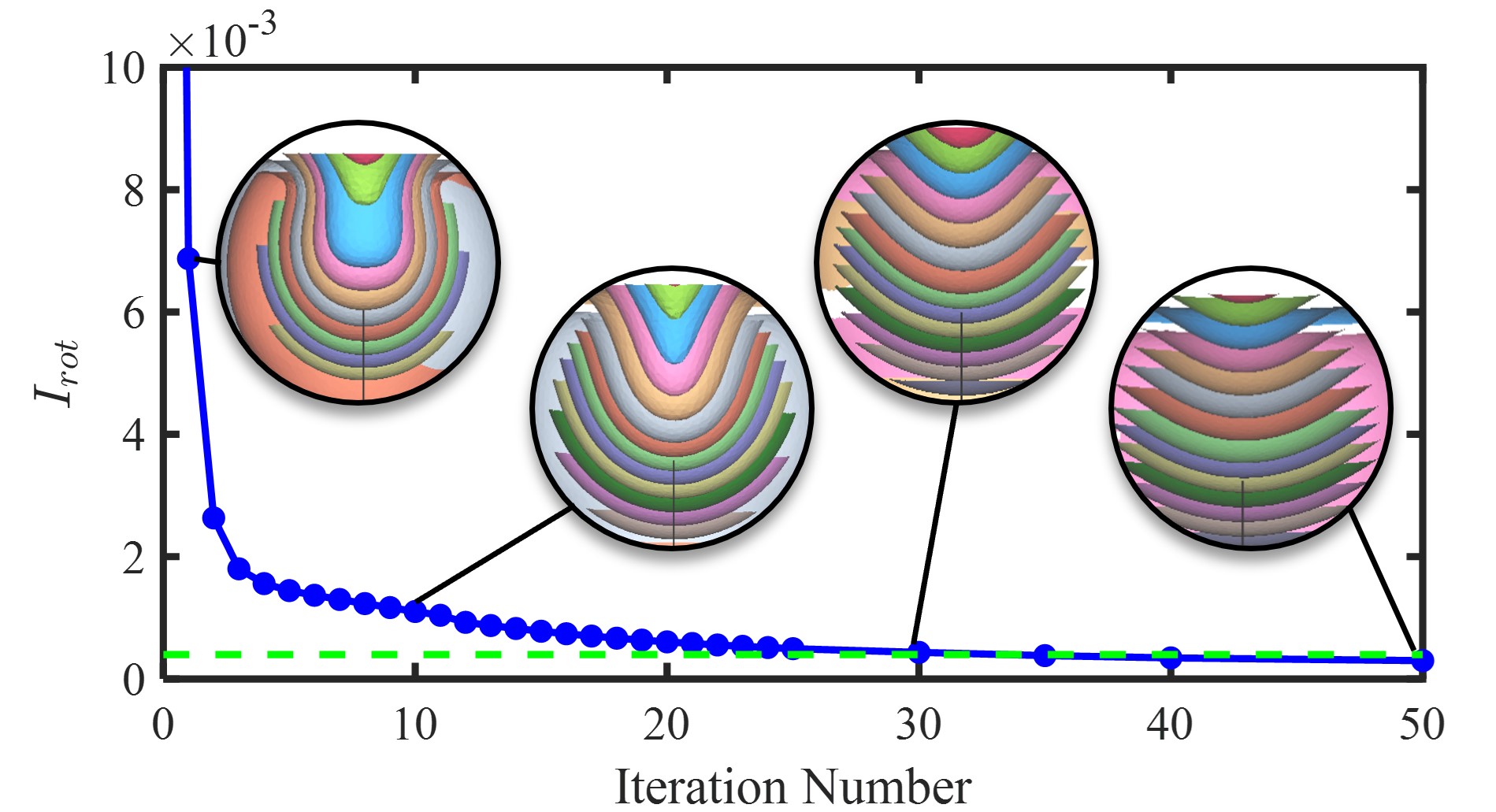}
\vspace{-16pt}
\caption{The energy-decreasing figure of $I_{rot}$ which evaluates the likelihood to be an irrotational vector field. The green line indicates our threshold of $4 \times 10^{-4}$ used for the terminal condition.
}\label{fig:iterationGraph}
\end{figure}

As shown in Fig.~\ref{fig:isolatedSingularity}, the method presented in Sec.~\ref{subSecSingularityResol} can successfully correct the Type-I singularities. However, around the region of singularity (e.g., center of the cup in Fig.~\ref{fig:isolatedSingularity}(b)), the layers are distorted after correction. We can find that the distances between the layers are not uniform. Upon checking the average value of  $I_{rot}$, it was found to be 0.0465 -- i.e., significantly higher than our threshold. We then apply the method presented in Sec.~\ref{subSecCurlRemov} to it. Figure ~\ref{fig:iterationGraph} shows the error after several iterations of the algorithm, and it can show that the distances between layers become more and more uniform.

\subsection{Physical experiment}
\label{subsecPhysicalExp}
\begin{figure}[t]
\centering 
\includegraphics[width=\linewidth]{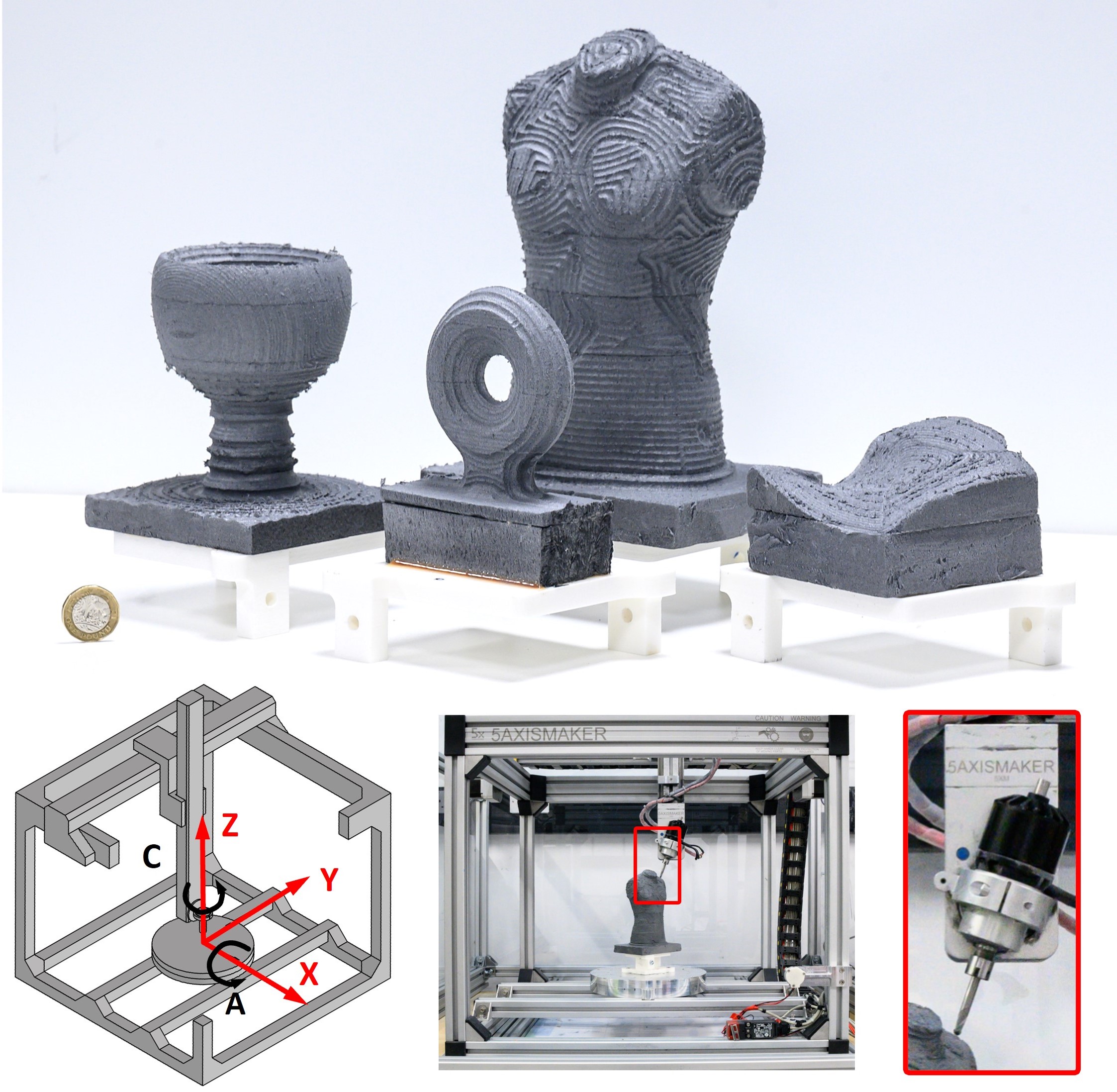}
\put(-254,100){\small \color{black}(a)}
\put(-254,5){\small \color{black}(b)}
\put(-172,5){\small \color{black}(c)}
\caption{The results of physical fabrication -- roughing conducted on example models with a \textsterling 1 coin placed for providing the scale reference (a). \revision{}{(b) CAD model of the machine used for physical fabrication. (c) The machine employed in our experiment is a modified 5XM600XL machine as described in \url{https://5axismaker.co.uk/5xm-model-comparison}. It has 3 linear axes, 1 rotational axis (A/B) on the head and 1 rotational axis (C) on the table. The working volume is 600mm x 600mm x 600mm. As the limit of B-axis is based on the length of cable, we restrict it to $\pm110$ degrees. The mechanical resolution is 0.036mm for the linear axes and 0.034 degrees for the rotary axes. A zoom-view of the tool is also shown in the right.}
}
\label{fig:resFab_all}
\vspace{10pt}
\end{figure}

In order to verify our methods, we conduct milling tests for different models processed through our method. A desktop level 5-axis Computer Numerical Control(\textit{CNC}) machine from the 5-Axis Works limited is used (see Fig.~\ref{fig:resFab_all}). We use a 6mm ball-end tool for all the operations. Figure~\ref{fig:layerBasedProcPlan}(c) shows the result of the roughing operation on a mannequin model. Figure~\ref{fig:resFreeform}(second row) shows the result of roughing corresponding to layers shown in Fig.~\ref{fig:resFreeform} (first row). We can see that using our method, the shape of the layers can be controlled so that a uniform layer of material is encountered during finishing (Fig.~\ref{fig:resFreeform}(last row)). This fact is also highlighted by Fig.~\ref{fig:resCmp_cup}, which compares the cutting depth variation encountered (computed) during finishing the parts obtained through ours and the traditional roughing strategies.  Using the comparison in Fig.\ref{fig:resFreeform}, we also highlight the flexibility of our approach as all three types of peeling is performed using our pipeline, depending on the requirement set by the user. Figure~\ref{fig:pipeline}(f) shows the roughed cup model using the layers shown in Fig.~\ref{fig:pipeline}(e), which are the result of increasing accessibility in the layers generated from convex hull (the center part of Fig.~\ref{fig:pipeline}(b)). Figure~\ref{fig:resFab_all} which contains all results for different models, also shows the result of the torus model in which the problem of singularity is resolved (Fig.~\ref{fig:resCmp_dryer}). \revision{}{All the physical experiments of milling are conducted on the same machine. However, the method is not restricted to this specific type of machine and can be used for machines in other configurations by introducing appropriate anchor vectors. The example shown in Fig.~\ref{fig:pipeline} has demonstrated how the vector field can be modified to restrict high tool inclination near the platform.} 



\subsection{Discussion}
\label{subSecDiscussion}
\subsubsection{Relation with other methods}
To our knowledge, our method is the first to explicitly use vector fields to control the volume peeling for rough machining. Since a vector field implicitly represents a gradient scalar field, our pipeline can achieve results obtained through most of the scalar field-based methods, specifically the propagation field-based ones. We have already shown the results in which the vector field initiates from a convex hull surface, which is similar to the method presented by \cite{he2021geodesic}. Moreover, in \cite{he2021geodesic}, the curvature of the scalar field near the model surface has been controlled. We can achieve a similar effect too by making the following modification to the vector field obtained from the convex hull (in the neighborhood of the model boundary):

\begin{equation}
    \mathbf{v} = \alpha  \mathbf{v}_{CH} + (1-\alpha ) \mathbf{v}_{N}
    \label{heat method}
\end{equation}
Where $\mathbf{v}_{N}$ is the surface normal of the corresponding model face and $\mathbf{v}_{CH}$ is the original field. The parameter $\alpha \in [0,1]$ can be adjusted to get the desired curvature. Fig.~\ref{fig:pipeline}(d) shows the curved layers inside the cavity of the cup. 

Similarly, initiating the vector field from the model's boundary surface gives offset layers that can be self-intersection free. Planar peeling also becomes achievable. This provides an opportunity to enjoy the advantages of different roughing strategies when necessary. Moreover, the advantage of our pipeline is its extra flexibility introduced to the whole process by allowing the local control of the layers' shape to satisfy the  conditions of manufacturability. This is especially handy while working in a dynamic environment in which the conditions of accessibility can change frequently.

\subsubsection{Weight tuning}\label{subsubsecWeightTune}
The weights ($\alpha$ and $\beta_i$s) in Eq.\eqref{eq:optimisation} are responsible for controlling the distribution of the vector field in the whole domain. As already mentioned, the term associated with $\alpha$ is responsible to ensure that the distribution is as uniform as possible. Whereas, the term associated with $\beta$ is responsible for enforcing the constraints. In case, Eq.\eqref{eq:linear} has a solution, the weights will have no influence over it. This situation will occur when there is no constraint (in which case, there is a zero vector field overall domain) or when all the constraints enforce the same direction. In all other cases, the weights will decide the nature of the distribution. The effect of weight is prominently visible when two or more constraints are close and point in different directions which counter the smoothness/uniformity condition. When $\alpha$ has a very high magnitude relative to all the $\beta$ (say tends to infinity), then the vector field distribution relaxes the constraint directions and a uniform field is obtained. On the other extreme, a sharp variation is observed near the constraints. For our case, we use ${\beta}/{\alpha} = 10^5$ for all the general cases. In case of a very important constraint, which needs to be preserved (say singularity avoidance), we use ${\beta}/{\alpha} = 10^8$ which creates a hard-constraint-like situation. Our current assignment of the weights is based on a visual analysis of shapes generated. Ideally, it should depend on the location and type of constraints.

\revision{}{
\subsubsection{Dependence on mesh quality} 
The computation of our approach relies on volume meshes with good quality. Coarser meshes introduce more discretization errors. First, the detailed features of the model cannot be well captured. Second, the assumption of element-based linearization will lead to large error when computing the optimized fields and extracting the resultant surfaces. Additionally, the constraints (i.e., those anchor vectors defined in Eq.~\eqref{eq:optimisation}) may not be limited to a local region on coarse meshes. 
An irregular mesh can also lead to negative weights on a large number of edges in the cotangent-based Laplacian formulation, which can affect the stability of linear systems (ref.~\cite{alexa2020properties}). 
Fine and regular meshes are needed. 
However, the cost of computing time and memory increases significantly with respect to the number of elements. All the objectives can be met by using regular tetrahedral meshes with edge-length being around 1.5mm in our experiments. Statistics for the computing time and the mesh sizes can be found in Table~\ref{tab:comprarison}.
}

\subsubsection{Limitations}
During the whole of our pipeline, we constrain the vectors to be of unit magnitude. This allows us to obtain uniform depth between the layers as well as to control the shape of the layers. Allowing the magnitude to vary provides an extra degree of freedom but makes it difficult to control the quality of the vector field as the magnitude can contribute to parameters like curl as well. 
\revision{But it has not been a focus of our current implementation.}{But it was not a focus of our implementation.}

Our current implementation does not explicitly include a method to improve the manufacturability of layers generated when the singularity is in the form of a curve satisfying the boundary requirements. This can be encountered when two or more streams of vector fields merge together as in the case of intersecting holes. Since the layer formed in such a case contains points that do not satisfy the strong accessibility condition, further processing is required to modify the approach direction for those points. However, a volume segmentation method can be integrated (as in \cite{mahdavi2020vdac}) with our method to split the domain into multiple domains, separating each stream of vectors, thereby removing such singularity.
The implementation in our current work does not explicitly incorporate machine-information like workspace limitations or tool-geometry. All experimental tests in our work are conducted by ball-end milling. However, users can adjust the anchor directions to make the layers that satisfy the requirements of the factors such as tool-length. Note that the required \revision{}{total} tool-length is in fact influenced by both the peeling patterns as well as the shape of the target model \revision{}{when considering the accessibility}. 

\revision{}{By maintaining a nearly-constant tool inclination w.r.t. the surface and uniformly spaced tool-path, we assumed that a uniform depth-of-cut translates to a constant tool engagement away from the boundaries. In the future, we would like to adjust the layer shape based on more direct dynamic consideration. A requirement on the layer shape can be represented as a corresponding requirement on the direction of the vectors in that area. Additionally, the magnitude of the vectors can be varied to create a non-uniform layer thickness. However, it is a research problem about how to control the  magnitudes effectively while preserving the desired quality of vector fields.}

\revision{Also, the}{The} current method does not emphasize much on the automation of seed / anchor point or constraint identification, nor does it control the surface-area of the layers generated. 
A function for detecting the accessible directions will help reduce the effort made by users for introducing anchor vectors to remove singularities. 
Finally, although not a major limitation, the use of a mesh-based representation reduces some degrees of freedom that can be achieved through a more implicit representation.


\section{Conclusion}
This paper has presented a novel volume peeling method for multi-axis machining that utilizes vector fields to achieve greater flexibility and constraint satisfaction in comparison to traditional scalar field-based methods. Our method utilizes an optimization formulation to generate an initial vector field aligned with user-specified anchor vectors and further optimizes the field to be irrotational for easier realization by scalar field gradients. We employ iso-surfaces of the scalar field as the layers of working surfaces for multi-axis volume peeling and have developed algorithms to address singularities of the fields. Our method has been thoroughly tested and verified through physical experimental machining on a variety of models, demonstrating its potential to enhance multi-axis machining in manufacturing applications. 

\section*{Acknowledgement}
The project is financially supported by the chair professorship fund of the University of Manchester and the industrial fund of 5AXISWORKS Ltd via the Innovate UK Smart Grant.

\section*{Appendix}

\begin{wrapfigure}[4]{r}{0.30\linewidth}\vspace{-45pt}
\begin{center}
\hspace{-25pt}\includegraphics[width=1.1\linewidth]{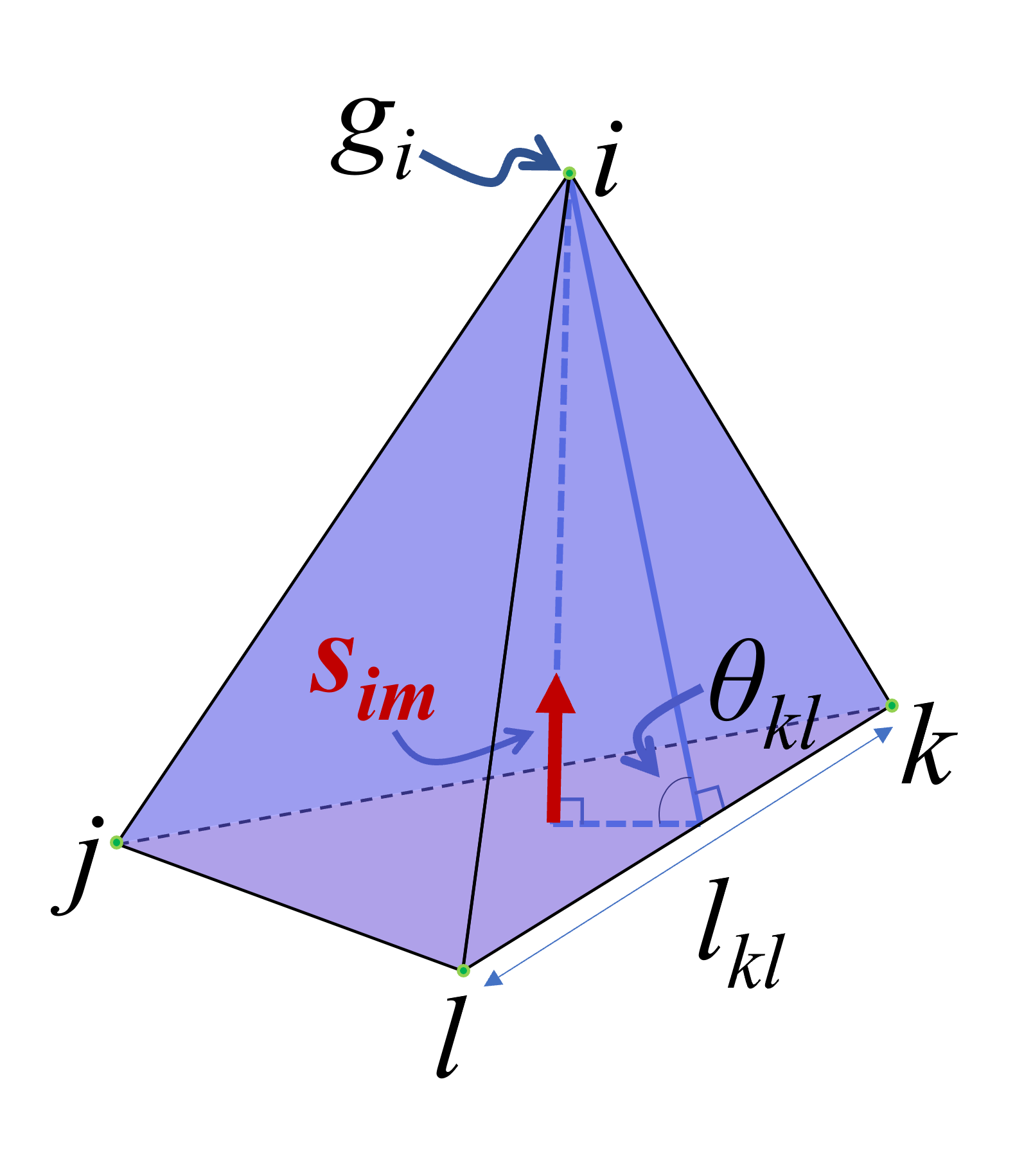}
\end{center}
\end{wrapfigure}


On a tetrahedral mesh, the discrete gradient is defined as shown in Eq.~\eqref{eq:gradient}. For a tetrahedron (represented in the adjacent figure), it is represented as \cite{liao2009gradient}:

\vspace{20pt}
\begin{equation}
\nabla g(t) = 
\begin{bmatrix}
       (\mathbf{x_i}-\mathbf{x_l})^T \\ 
       (\mathbf{x_j}-\mathbf{x_l})^T \\
       (\mathbf{x_k}-\mathbf{x_l})^T
\end{bmatrix}^{-1} \begin{bmatrix}
    1&0&0&-1 \\
    0&1&0&-1 \\
    0&0&1&-1 
\end{bmatrix}
\begin{bmatrix}
    g_i \\
    g_j \\
    g_k \\
    g_l
\end{bmatrix}
\label{eq:gradientmatrix}
\end{equation}

The terms $\mathbf{x_{\{i,j,k,l\}}}$ in Eq.~\eqref{eq:gradientmatrix} represent the position vectors of the vertices $\mathcal{V}_{\{i,j,k,l\}}$ with respect to a global coordinate system for the whole domain. 

The (integrated) divergence (at the $i^{th}$ vertex, $\mathcal{V}_i$) is similarly computed as:
\begin{equation}
    \nabla \cdot \mathbf{v}(\mathcal{V}_i) = \frac{1}{3}\sum_{\mathcal{T}_m \in \mathcal{N}(\mathcal{V}_i)} \mathbf{s_{im}} \cdot \mathbf{v}_m,
\end{equation}
where $\mathbf{s_{im}}$ is the area vector of the face opposite to vertex $\mathcal{V}_i$ in the tetrahedron $\mathcal{T}_m$. And $\mathcal{N}(\mathcal{V}_i)$ is the neighbourhood of vertex $\mathcal{V}_i$ including all tetrahedrons incident to it. This vector is used for the right-hand side of the Eq.~\ref{eq:discretePoisson} which is the integrated divergence.

The Co-tangent Laplacian matrix for a tetrahedral mesh is defined as:

\begin{equation}
    L_c{(i,j)} = 
    \begin{cases}
    -w_{ij} & \text{for }  \mathcal{V}_j \in \mathcal{N(V}_i) \text{, }i \neq j \\
    
     \sum_{\mathcal{V}_k \in \mathcal{N(V}_i)}^{} w_{ik}   & \text{for }i = j \\
    0  & \text{for }  \mathcal{V}_j \notin \mathcal{N(V}_i)
    
    \end{cases}
\end{equation}
where,
\begin{equation}
    w_{ij} = \frac{1}{6} \sum_{\mathcal{T}_m \in \mathcal{N(E}_{ij})} l_{kl} cot(\theta_{kl})
\end{equation}
The diagonal matrix $\mathbf{M}$ in Eq.~\eqref{eq:discretePoisson} is such that each $i^{th}$ diagonal element is equal to one-fourth of the total volume of all elements (tetrahedrons) incident to the $i^{th}$ node.

\section*{References}

\bibliography{main}

\end{document}